\documentclass[aps,prb,reprint,showpacs,superscriptaddress,nofootinbib,floatfix,noshowpacs,showkeys]{revtex4-1}

\usepackage{color}
\usepackage{amsmath}
\usepackage{amssymb}
\usepackage{graphicx}
\usepackage{bm}
\usepackage{tikz}

\DeclareMathOperator\erfc{erfc}
\DeclareMathOperator\erf{erf}

\newcommand{\braket}[2]{\langle #1|#2\rangle}
\begin{document}

\title{Self-consistent quasi-particle $GW$ and hybrid functional calculations for Al/InAs/Al heterojunctions:
band offset and spin-orbit coupling effects}

\author{H. Ness}
\affiliation{Department of Physics, Faculty of Natural and Mathematical Sciences,
King's College London, Strand, London WC2R 2LS, UK}

\author{F. Corsetti}
\affiliation{Microsoft Azure Quantum, 2800 Lyngby, Denmark}

\author{D. Pashov}
\affiliation{Department of Physics, Faculty of Natural and Mathematical Sciences,
King's College London, Strand, London WC2R 2LS, UK}

\author{B. Verstichel}
\affiliation{Synopsys Denmark, 2100 Copenhagen, Denmark}

\author{G. W. Winkler}
\affiliation{Microsoft Azure Quantum, Goleta, California 93111, USA}

\author{M. van Schilfgaarde}
\affiliation{Department of Physics, Faculty of Natural and Mathematical Sciences,
King's College London, Strand, London WC2R 2LS, UK}
\affiliation{National Renewable Energy Laboratory, Golden, Colorado 80401, USA}

\author{R. M. Lutchyn}
\affiliation{Microsoft Azure Quantum, Goleta, California 93111, USA}

\begin{abstract}
The electronic structure of surfaces and interfaces plays a key role in the properties of quantum devices. 
Here, we study the electronic structure of realistic Al/InAs/Al heterojunctions using a combination of density functional
theory (DFT) with hybrid functionals and state-of-the-art quasi-particle $GW$ (QS$GW$) calculations.
We find a good agreement between QS$GW$ calculations and hybrid functional calculations which themselves
compare favourably well with ARPES experiments.
Our study confirm the need of well controlled quality of the interfaces to obtain the needed properties of InAs/Al heterojunctions.
A detailed analysis of the effects of spin-orbit coupling on the spin-splitting of the electronic states show a linear scaling 
in $k$-space, related to the two-dimensional nature of some interface states.
The good agreement by QS$GW$ and hybrid functional calculations open the door towards trust-able use of an effective approximation to QS$GW$
for studying very large heterojunctions.
\end{abstract}

\keywords{electronic structure of surfaces and interfaces, density functional theory, hybrid functionals and many-body perturbation theory,
quantum materials, III-V semiconductors}

\pacs{???}

\maketitle

\section{Introduction}
\label{sec:intro}

Because of their unique combination of material parameters 
(i.e. large spin-orbit coupling, small effective mass, large Lande g-factor),
narrow-gap III-V semiconductors (such as InAs or InSb)
have generated considerable interest in many technological applications.

Recently, these materials have been central to the experimental 
realisation of the so-called Majorana zero modes
\cite{Sau:2010,Alicea:2010,Lutchyn:2010,Oreg:2010,Alicea:2012,Leijnse:2012,Beenakker:2013,DasSarma:2015,Aguado:2017,Lutchyn:2018}. 
In this devices, the main goal is to develop topological $p$-wave 
superconductivity at the interface of a conventional semiconductor and 
an $s$-wave superconductor. 
An exceptionally good control of the interface properties is needed to realise 
topological superconducting phases and to manipulate Majorana zero modes
which are the key ingredient in topological quantum computation proposals \cite{Nayak:2008,DasSarma:2015,Lutchyn:2018,parityInAsAl:2024}. 
The hybrid semiconductor-superconductor Majorana devices
are required to have a large g-factor, strong Rashba
spin-orbit coupling and significant proximity-induced 
superconducting gap. 
Recently, the proximity-induced superconductivity has been studied 
in devices made of a semiconductor nanowire in contact with a superconductor,
including 
Al/InAs \cite{Krogstrup:2015,Suominen:2017,Nichele:2017,Vaitiekenas:2018,Matsuo:2020,Menard:2020,Vaitiekenas:2020}, 
Al/InSb \cite{deMoor:2018,Shen:2018,Anselmetti:2019,OhVeld:2020}
Pb/InAs \cite{Kanne:2021}, and Sn/InSb \cite{Pendharkar:2021}. 
High-quality superconductor/semiconductor 
interfaces (i.e. uniform and transparent) are required to optimise the topological 
gaps in these heterostructures.

The geometry of the interface may give rise to
(desirable or undesirable) interface states, it may alter the band
bending and band alignment, or affect the magnitude of
the proximity-induced gap and of the spin-orbit coupling.
Understanding the resulting surface/interface states and 
Fermi-level pinning is important for engineering appropriate interface 
Hamiltonian and realising topological superconductivity hosting Majoranas.

Band bending and surface states have been observed by angle-resolved photoemission 
spectroscopy (ARPES)\cite{Tomaszewska:2015,Schuwalow:2021,SYang:2022} and scanning tunneling microscopy and spectroscopy (STM/STS)\cite{PhysRevMaterials.7.066201}.
First principles simulations based on density functional theory (DFT) can help 
interpret experiments and resolve the effects of the interfaces.
DFT studies of InAs and InSb surfaces and interfaces have been limited because 
local (local density approximation LDA) and semi-local exchange-correlation functionals severely underestimate
the band gap to the limiting point where it reduces to zero \cite{Krukau:2006}.
More accurate methods involving quasi-particle self-consistent 
$GW$ (QS\emph{GW}) approaches or hybrid functionals provide results much closer
to the experimental (bulk) gap (0.42 eV for bulk InAs).

In this paper, we present calculations of realistic Al(111)/InAs(001) 
heterojunctions using a QSGW method implemented in the Questaal package.
The QS\emph{GW} results are also compared with hybrid functionals DFT calculations.
We focus our attention on the effects of ``disorder'' (using numerical ``experiments'') 
on
the electronic structure of realistic InAs/Al interfaces described at the atomic scale.
The disorder we consider is coming from: (i) atomic relaxations (i.e. the atoms at
the InAs/Al interfaces do not rest at their correspond bulk atomic positions), (ii)
substitution disorder which mimics in a simple way potential atomic diffusion at
the interface, and (iii) rescaling the spin-orbit coupling strength on some atoms
which mimics the presence of some external electric fields at the interfaces.

The paper is organised as follows: In Section \ref{sec:calc}, we present the InAs/Al 
system we considered and the two software packages used for the electronic 
structure calculations, namely the Questaal and QuantumATK packages.
The results of our calculations are shown and analysed in Section \ref{sec:res}, 
where we extract the profiles of the valence band maxima (VBM) and conduction band minim (CBm) along
the InAs/Al heterojunction and study in details the effects of spin-orbit coupling
(SOC) on some specific bands. Conclusions are presented in Section \ref{sec:ccl}.
Additional informations are provided in the appendices, about: the implementation 
of the Questaal code on GPUs in Appendix \ref{app:gpu}, the hybrid functional in Appendix \ref{app:ddh}, local density
of states, and bulk versus heterojunctions bands in subsequent appendices.

\section{Calculations}
\label{sec:calc}

First principles electronic structure calculations have been performed using two different packages:
the Questaal package \cite{questaal:web} and the pseudopotential QuantumATK package \cite{Smidstrup_2020}.

Questaal is an all-electron method, with an augmented wave basis consisting of partial waves inside augmentation
spheres based on the linear muffin-tin orbital (LMTO) technique \cite{Pashov:2020}.
It includes conventional DFT-based calculations, as well as many-body perturbation theory, especially with its implementation 
of a quasi-particle self-consistent \emph{GW} (QS\emph{GW})
approach \cite{Faleev:2004,Kotani:2007}.

\begin{figure}
\centering
\includegraphics[width=25mm]{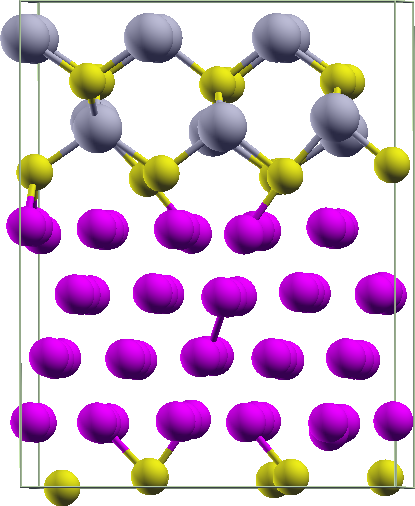}\includegraphics[width=25mm]{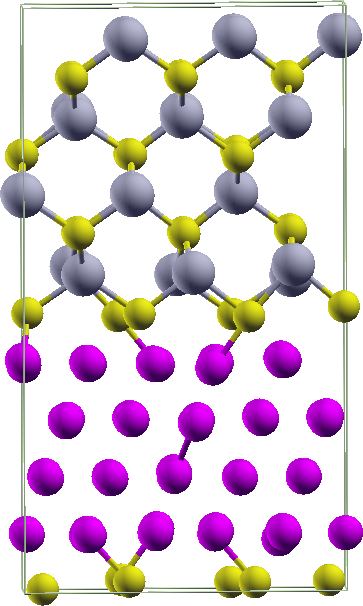}\includegraphics[width=25mm]{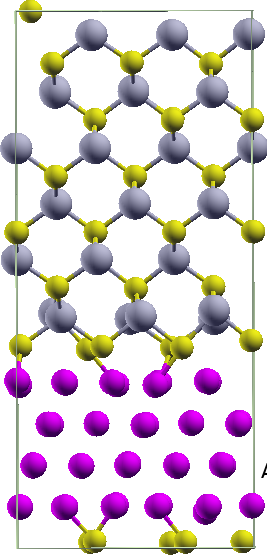}\\
\vskip 5mm
\includegraphics[width=60mm]{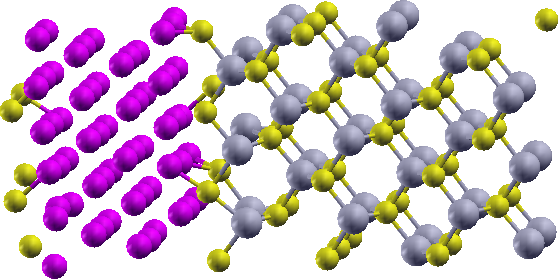}
\caption{\label{fig:atpos}
(Top) Ball-and-stick representation of the relaxed InAs(001)/Al(111) supercell, for different number of InAs layers. In atoms are shown in grey, As atoms in yellow and Al atoms in purple. (Bottom) Side view of the InAs(001)/Al(111) atomic planes. The largest supercell contains 138 atoms with 4, 6, 7 atomic planes of In, As, Al respectively. Each plane contains either 6 atoms of In (As) or 15 atoms of Al.
}
\end{figure}

We have considered InAs/Al heterojunctions for which the interface between the two materials
is built from the (001) surface for InAs and from the (111) surface for Al, with As-terminated
InAs surfaces in direct contact with the Al surface (as suggested by the experiments in \cite{Schuwalow:2021}).
We took a low temperature lattice parameter of  $a_0=6.06$ \AA\ for InAl, and
a (001) surface supercell based on the two following (001) surface vectors $u_1=[2,0]a_0$
and $u_2=[-1,3]a_0$. For this (001) supercell, there are 6 atoms
in each In (As) atomic planes perpendicular to the $z$-direction, see Fig.~\ref{fig:atpos}.
For this supercell, one can match the Al(111) surface rather well, with a slight stretch
(of 3\%) in the $u_2$ direction, using a bulk lattice parameter of $a_\text{Al}=4.05$ \AA.
Then, each Al atomic plane parallel to the InAs/Al interface contains 15 atoms.

In order to minimise the computational cost, more specifically for the QS\emph{GW} calculations,
we have considered the minimal possible size for the junctions.
We have found that to be able to keep the bulk-like character for the electronic structure in the middle of 
the InAs slab, one needs to go beyond a few layers of InAl: typically for 6 (7) atomic planes of In (As) (and beyond)
we recover the bulk-like density of states for the In (As) atoms in the center of the InAs slab.
As Al is a metal with a shorter screening length, fewer atomic layers are needed (typically 4 atomic (111) planes
are enough) to obtain a bulk-like density of states in the central atomic layers.

Relaxation of the atomic positions have been performed within DFT-LDA. We have allowed the atoms in the Al atomic layers next 
to the interfaces, and the atoms of As and In in the two outmost atomic layers close to the InAs/Al interfaces to relax until 
the force components are below 10 mRy/bohr (257 meV/\AA). 

We did not impose any symmetry during the atomic relaxation. Therefore the two InAs/Al interfaces of the supercell are not 
equivalent. This allows us to minimise the possible existence of unwanted electronic states that may have arisen due to size 
and coupling effects between two perfectly symmetric interfaces. To some extend, this can be also seen as a simplified case of 
``geometric'' disorder.

In DFT-LDA and even in $GW$ calculations, band gaps are often underestimated (LDA) or overestimated ($GW$).
Indeed they should be because the RPA screened Coulomb interaction $W$ is not sufficiently screened. 
Improvement of $W$ by the addition of ladder diagrams indeed does improve the gaps \cite{Cunningham:2023,Kutepov:2017}. 
In most of the cases, the dielectric constant of semiconductors and insulators is about 80\% of the experimental value. 
This is because the ladder diagrams are missing in the RPA. 
However, these higher order diagrams are computationally costly, and here we adopt a simpler approach here. 
We have found that scaling the dielectric constant by 0.8, or alternatively using a hybrid of 80\% QS$GW$ and 20\% LDA, 
we can mimic the effect of the ladders. This eliminates most of the errors.

Hence, we have use hybrid of LDA and QS$GW$ functionals,
$\Sigma^{\rm scaled}=\Sigma_{\rm QSGW} \times 0.8 + V_{\rm xc}^{\rm LDA} \times 0.2$,
in the calculations of our band structures.

The system we considered contains 138 atoms ($4\times 15$ atoms of Al, $6 \times 6$ of In and $7 \times 6$ of As).
To our knowledge, this is one of the first time that
self-consistent QS\emph{GW} calculations have been performed for such a ``large'' system. 
The Questaal package was re-developed and optimised to take advantage of GPU-based computing on a multi-petaflop modular 
supercomputer, see Appendix \ref{app:gpu}.

We are interested in determining the profile of the valence band maxima (VBM) and of the conduction band minima (CBm)
along the InAs/Al heterojunction.
There are different ways to find such a profile, for example by considering the electrostatic potential of the heterojunction,
or by considering the change in energy position of deep electronic levels of the junction in comparison to their bulk
equivalent. Obviously, the profile of the VBM or CBm obtained from such atomic scale systems (for example, the thickness
of the InAs slab in the 138 atom supercell is $\sim 26.5$ \AA) 
will not reflect the band bending of Schottky barriers expected from a continuum model of the semiconductor/metal contact
described on the micron scale.
However our atomic scale calculations incorporate the more local effects of the InAs/Al interfaces against the bulk 
property of the materials.

To evaluate the profile of the valence (conduction) band maxima (minima), 
we extract the energy position of the deep electronic levels on each In atom (deep $d$-orbital) and each As atom
(deep $s$-orbital). Assuming a rigid energy shift of these deep electronic levels relative to the Fermi level $E_\text{F}$
in both the bulk and the heterojunction systems, we can determine the profile (averaged over the number of atoms in
each atomic layer) of the VBM (along the main direction of the junction) relative to the exact QS$GW$ Fermi
level of the junction. The profile of the CBm is obtained from a rigid shift of the VBM by the bulk QS$GW$
band gap (0.47~eV in the present case).

{It is important to note that all the QS$GW$ calculations were performed in the presence of spin-orbit coupling (SOC).
Orbitals with $s$-like character are not affected by the presence of SOC. However the $d$-orbitals are split
by the SOC. For bulk-like environment, the $d$-orbitals are split into two subsets according to the crystal symmetry.
For In atoms close the Al/InAs interfaces the symmetry is reduced (further reduced by the atomic relaxations), and different
energy shifts occur for the different $d$-orbitals on these atoms. There is more ``fluctuation'' of the energy shift
for these atoms in comparison to bulk-like In atoms in the center of the InAs slab.}

{As a final comparison, we also perform calculations on the same system using a different methodology for correcting the band gap problem: hybrid functionals. Traditional hybrid functionals which use a fixed global mixing fraction $\alpha$ of Fock exchange can show limitations in the case of inhomogeneous interface systems, specifically if the different materials require different values of $\alpha$ to recover the correct bulk electronic structure and band gap. The situation is even more severe for the present case of semiconductor/metal interfaces.

To overcome this problem, we build on a recently developed scheme for a local (i.e., spatially-varying) mixing fraction, based on an estimator of the local dielectric constant defined as a functional of the electronic density \cite{PhysRevB.83.035119, doi:10.1021/acs.jctc.7b00853}. In order to deal with the metallic region in our system, a second, metallic estimator, is introduced, which determines locally if the material is a metal, and, if so, sets $\alpha$ to zero. A more detailed explanation of the method can be found in Appendix \ref{app:ddh}. We apply this scheme to the HSE06 functional \cite{10.1063/1.2404663}. We shall refer to these calculations as HSE06+DDH (dielectric-dependent hybrid).

The HSE06+DDH calculations are performed using the QuantumATK package \cite{Smidstrup_2020} (version T-2022.03) within a pseudopotential and 
linear combination of atomic orbitals (LCAO) formalism. The calculations are carried out with a spin-polarized non-collinear Hamiltonian. 
We use norm-conserving pseudopotentials from the PseudoDojo \cite{VANSETTEN201839} fully relativistic set and the medium basis 
set (LCAO-M\cite{Smidstrup_2020}) from QuantumATK. 
The auxiliary density matrix method (ADMM) \cite{doi:10.1021/ct1002225} is used to speed up the calculation of the exchange matrix.

The determination of the VBM/CBm is performed in the same way as for the QS$GW$ calculations, i.e. by extracting the energy position of 
the core levels relative  to $E_\text{F}$. 
However, due to the configuration of the pseudopotentials, we use the semi-core $d$-orbitals for both In and As. The bulk InAs band gap obtained 
with HSE06+DDH is 0.47~eV, in agreement with QS$GW$.}

\section{Results}
\label{sec:res}

\subsection{Band alignment}
\label{sec:vbmcbm}

\begin{figure}
\centering
\includegraphics[width=80mm]{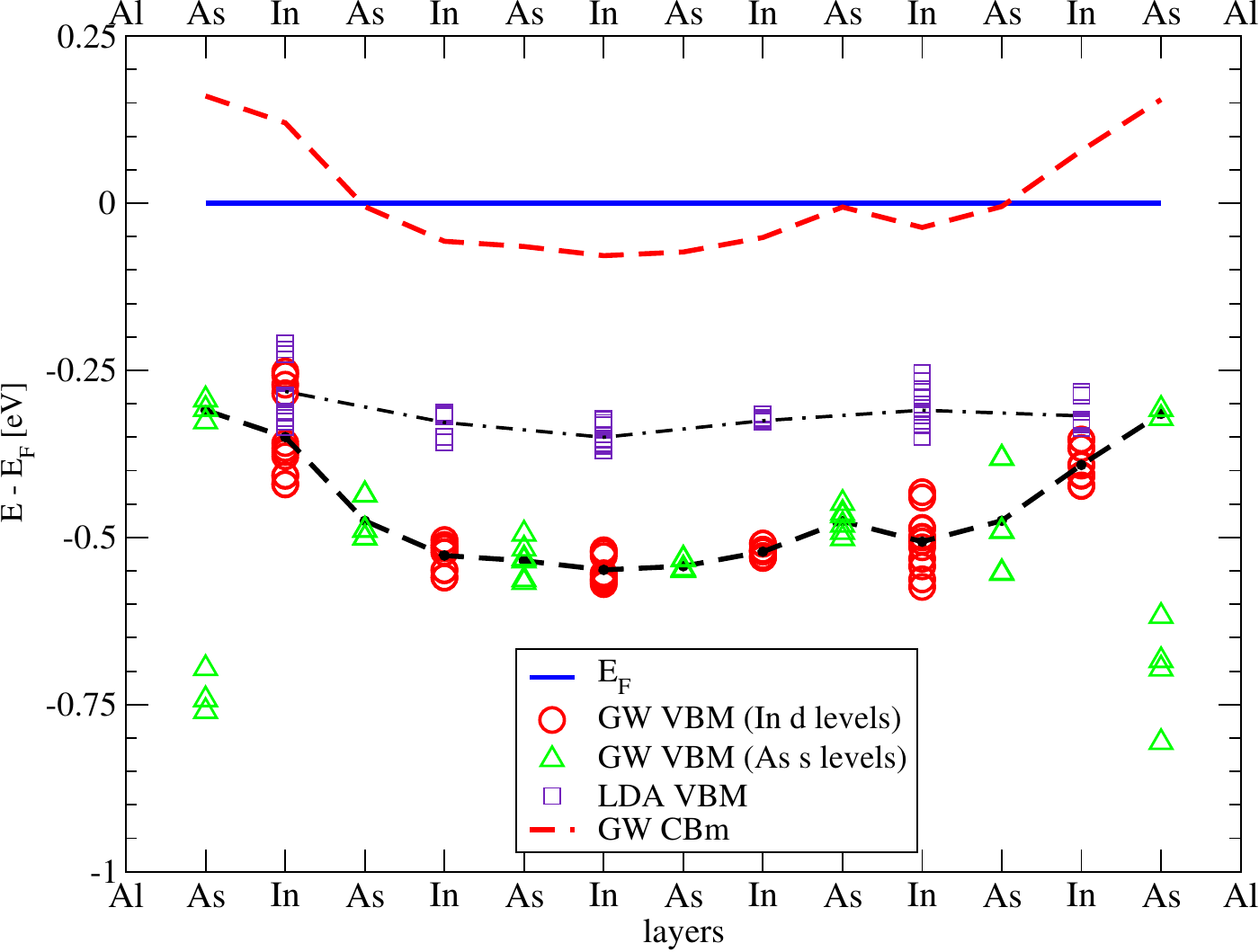}
\caption{\label{fig:localVBM}
Profile of the local valence band maximum (VBM) for the InAs/Al junction made of 138
atoms.
Symbols represent the different shifted energy levels (deep In-$d$ and As-$s$ orbitals) 
of the In and As atoms located in each atomic plane along the main axis of the junction.
Lines correspond to the averaged value of the energy shift in each atomic plane.
The black dashed and dashed-dotted lines correspond to the QS$GW$ and LDA VBM respectively. 
The red dashed line is the QS$GW$ conduction band minima (CBm), i.e. the VBM shifted
by the QS$GW$ bulk band gap.
Blue line presents the Fermi level $E_\text{F}$ position. Both QS$GW$ and LDA have been
shifted to correspond to this energy reference $E_\text{F}=0$. 
The QS$GW$ CBm lies $\sim$80 meV below $E_\text{F}$ in the center of the InAs slab.
}
\label{fig:qsgwldaVBM}
\end{figure}

Figure \ref{fig:localVBM} shows the profile of the VBM in the InAl/Al heterojunctions. It corresponds to a shallow 
parabolic-like curve where the VBM is higher close to the
InAs/Al interfaces than in the bulk-like part of InAs. Such a profile does not directly compare with conventional
band-bending in Schottky barriers, the latter occurs on much larger length scales ($\sim \mu\text{m}$) than
the scale corresponding to our atomic-scale calculations.
However, the inflections of the VBM reflects the effect of the interfaces against the bulk, most
certainly due to the presence of interface electric dipoles.

The dispersion in the energy shifts displayed by the symbols (red circles for the In deep $d$-orbital, green up-triangles for the As deep $s$-orbital) reflects that each atom in a given atomic layer are different.
This is mostly true for the atoms near the InAs/Al interfaces which have been allowed to relax. Such a
dispersion is minimal for the atoms in the center of the InAs slab where their position correspond to bulk
unrelaxed atomic positions.

The QS$GW$ calculations for bulk InAs provides a band gap 0.47~eV, close to the experimental band gap of 0.42~eV at low temperature. 
From that value, we can see 
that the CBm (obtained from rigid shift of the VBM) crosses the Fermi level of the junction. 
The CBm lies below $E_\text{F}$, by an amount of $\sim$80 meV, in the ``bulk'' part of the InAs slab. 

\begin{figure}
\centering
\includegraphics[width=80mm]{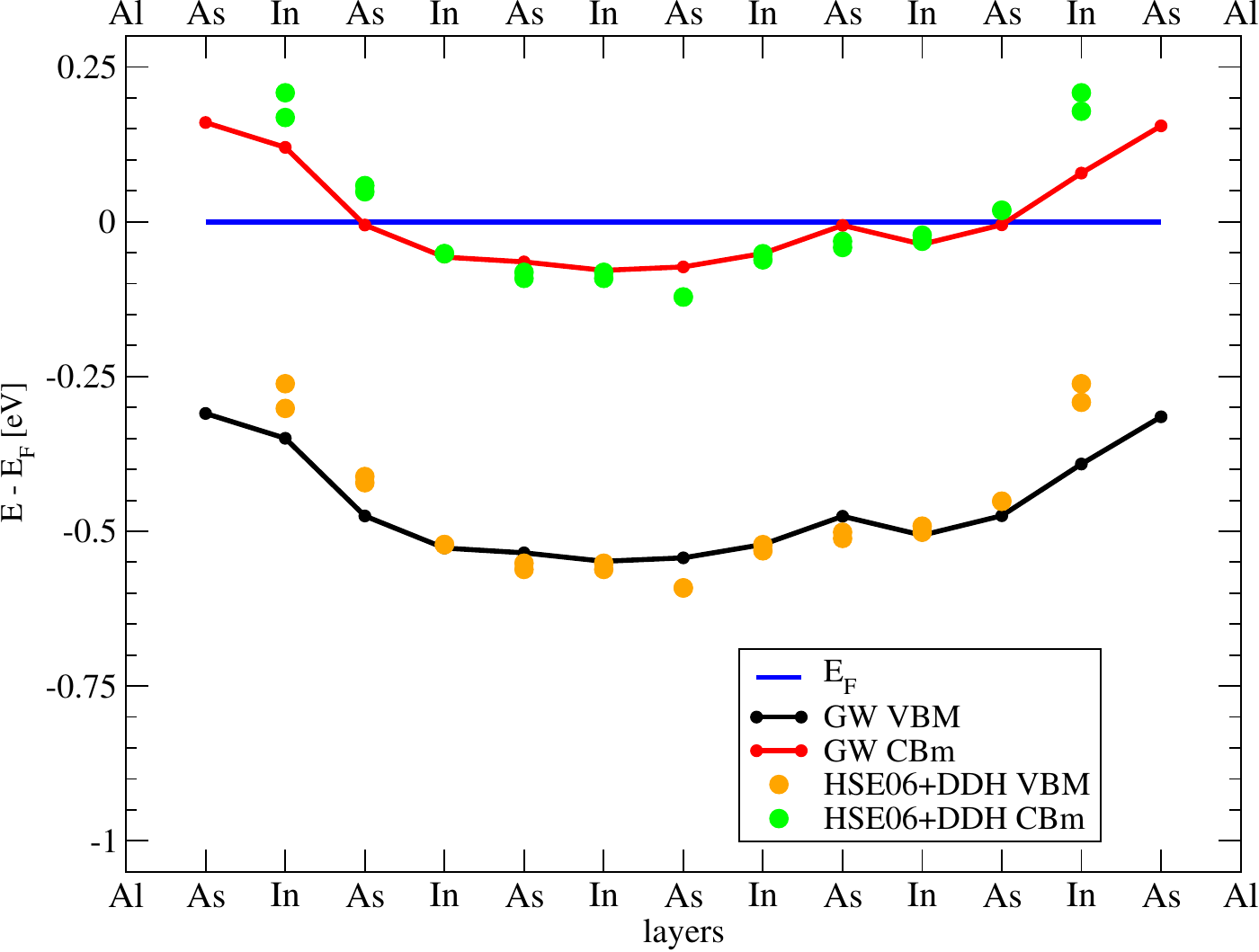}
\caption{\label{fig:localVBM_hse_gw}
Profile of the local VBM and CBm for the InAs/Al junction obtained from QS$GW$ (same values as in Figure~\ref{fig:localVBM}) and HSE06+DDH. 
The agreement between the two methods is particularly good.
}
\end{figure}

{This prediction of an accumulation layer in InAs is in agreement with ARPES measurements \cite{Schuwalow:2021}, although the band offset extracted 
from experiment is larger than what is seen in our simulation cell. For quantitative agreement, convergence to larger cells is probably needed.

For this, we turn to our hybrid functional calculations using the HSE06+DDH method. This method represents a more empirical and less accurate approach than QS$GW$, but also significantly less demanding of computational resources, and therefore potentially able to scale to larger systems. Figure~\ref{fig:localVBM_hse_gw} shows the comparison of the local band edges calculated with QS$GW$ and HSE06+DDH. The overall shape of the band edges is well reproduced by HSE06+DDH, and there is an excellent quantitative agreement in the bulk of the semiconductor (within two layers of the interface). As an additional check, we have also compared the LDA band edges calculated with the package Questaal (shown in Figure \ref{fig:localVBM}) with QuantumATK (not shown) and recovered a similar good agreement. 
Therefore, the HSE06+DDH method provides an effective approximation to the full QS$GW$ for this system, and might be used to explore much larger interface cells 
or multiple different interface configurations at a lower computational cost.}

In order to get a better understanding of the effects of disorder at the InAs/Al interfaces
on the energy level alignment, we have considered the following ``numerical experiment''.

We simulate the possible atomic diffusion at the interfaces by substituting some atoms by others.
From the valence properties of In, As and Al, it appears reasonable to envisage substitutional disorder
between In and Al atoms. We have therefore considered InAs/Al interfaces for which some In atoms 
(in the atomic layers closest to the interfaces) are replaced by Al atoms.

\begin{figure}
\centering
\includegraphics[width=80mm]{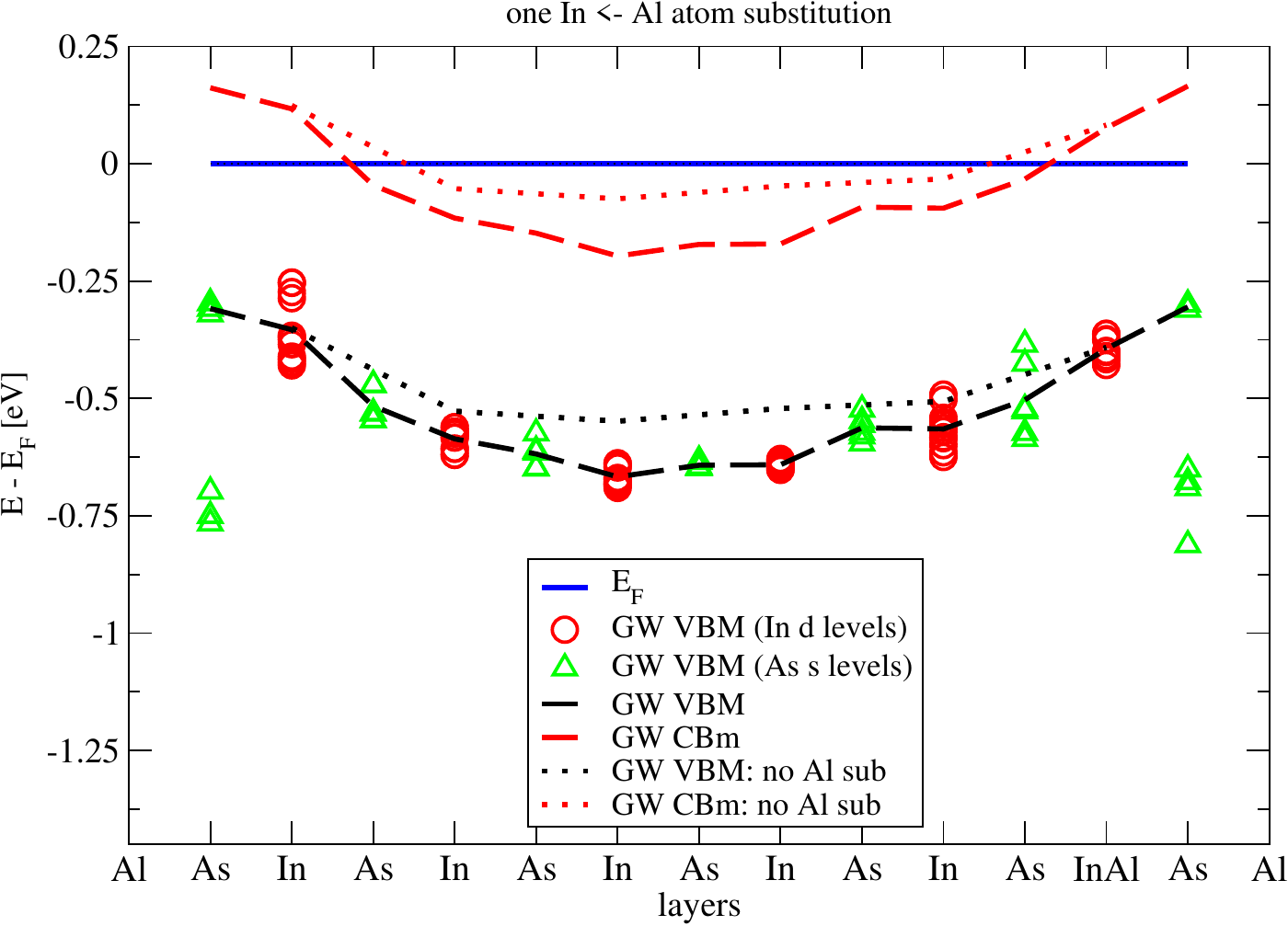}
\caption{\label{fig:localVBM_Al1a}
Profile of the local valence band maximum (VBM) for the InAs/Al junction with
similar notations as in Fig \ref{fig:qsgwldaVBM}.
Comparison between perfect and ``disordered'' interfaces. One atom of In is replaced by
one Al atom in the In atomic plane closest to the right InAs/Al interface (see label
InAl on the horizontal axis). All calculations are performed with QS$GW$.
The VBM/CBm are pushed down to lower energy in the case of the ``dirty'' interface, 
i.e. compare dashed-lines with dotted-lines (black for VBM, red for CBm).
``Interface disorder'' seems to push the CBm down by a further $\sim$ 100 meV.
A similar behaviour has also been obtained in two different cases of one In $\leftarrow$ Al
atom substitution (the Al atom is located at a different site in the corresponding In atomic
plane), as well as for a case of 2 In $\leftarrow$ Al
atoms substitution in the corresponding In atomic plane.
}
\end{figure}

There are many possible combinations to realise such substitutions, and we have considered only a
few of them. We started by replacing only one In atom by one Al atom in the In atomic layer located
the closest to the the right InAs/Al interface, see labels for the atomic planes
in Figure~\ref{fig:localVBM_Al1a}. 
We have performed calculations for only two cases over the six possible cases of one atom substitution. 
The results for the VBM profile for one of this case is shown in Figure~\ref{fig:localVBM_Al1a}.
We have also considered one case in which two In atoms are swapped by two Al atoms and we have
found similar trends for the profile of the VBM.
Our calculations indicate that the VBM is pushed down, by a further $\sim$100 meV, to lower energy 
in the case of the ``dirty'' interfaces compared to the case of ``perfect'' interfaces.

\subsection{Spin-orbit coupling effects}
\label{sec:sclsoc}

The presence of Rashba-like spin-orbit coupling (SOC) in narrow-gap InAs semiconductors is one of the central
ingredients for inducing superconducting property by proximity of an $s$-wave superconductor like Al. 
Once superconducting, a InAs nanowire can eventually hosts a pair of Majorana state at each of its ends
where the superconducting order parameter vanishes.

In the previous section, we have studied how band alignment in InAs deviates from pure bulk to InAs/Al
interfaces, including some form of disorder of the interfaces.

We now consider the possibility of another type of disorder and its effects on the band structure of the InAs/Al junction.
For that we now consider the following numerical ``experiment'': the strength of the SOC, on some atoms in the system,
is rescaled to larger values. The SOC rescaling is applied on either all the In and As atoms or
only on the In and As atoms close to the InAs/Al interfaces (i.e. In atoms labelled $z=12,z=27$ and As atoms labelled
$z=0.3,z=10$ in panel (d) for Fig.~\ref{fig:localDOS}). Note that a light element like Al does not have strong
SOC and rescaling the SOC on Al is not relevant.
The increase of the SOC can be seen as an indirect effect of the presence of an extra external electric field (perpendicular
to the InAs/Al interface in the case of Rashba-like SOC) due to gating or other effects  
not taken into account in our model of the InAs/Al heterojunction.

In Appendix \ref{app:bnds}, we show how the band structure of the InAs/Al junction differs strongly from the bulk InAs
bands due to the coupling to the metallic As states. Most of the bands in the junction come from a mixture of all
In, As and Al orbitals.
However, we have identified an equivalent to the bottom of the bulk InAs conduction band for the
case of the heterojunction. And we calculate how the energy difference $\Delta E(k)$ between the SOC spin-split bands 
(around the $\Gamma$ point) varies with the rescaling to the SOC.

\begin{figure}
\centering
\text{(a)}\includegraphics[width=35mm]{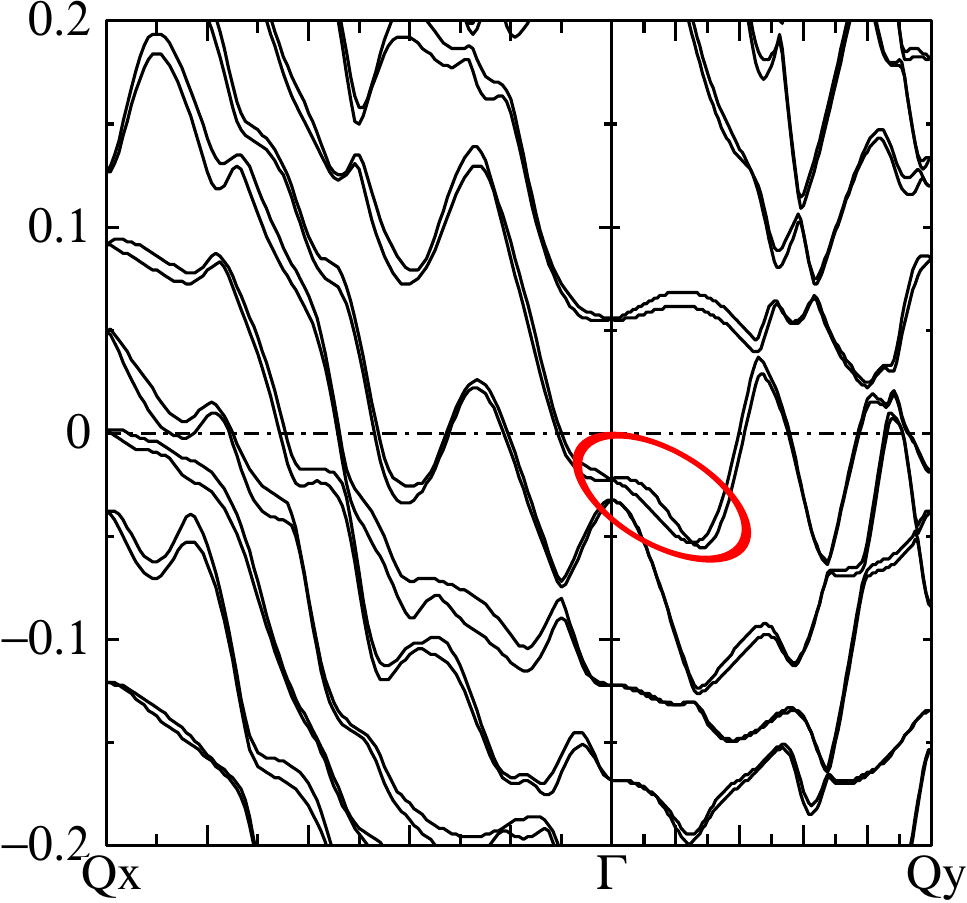}\hspace{5mm}\text{(b)}\includegraphics[width=35mm]{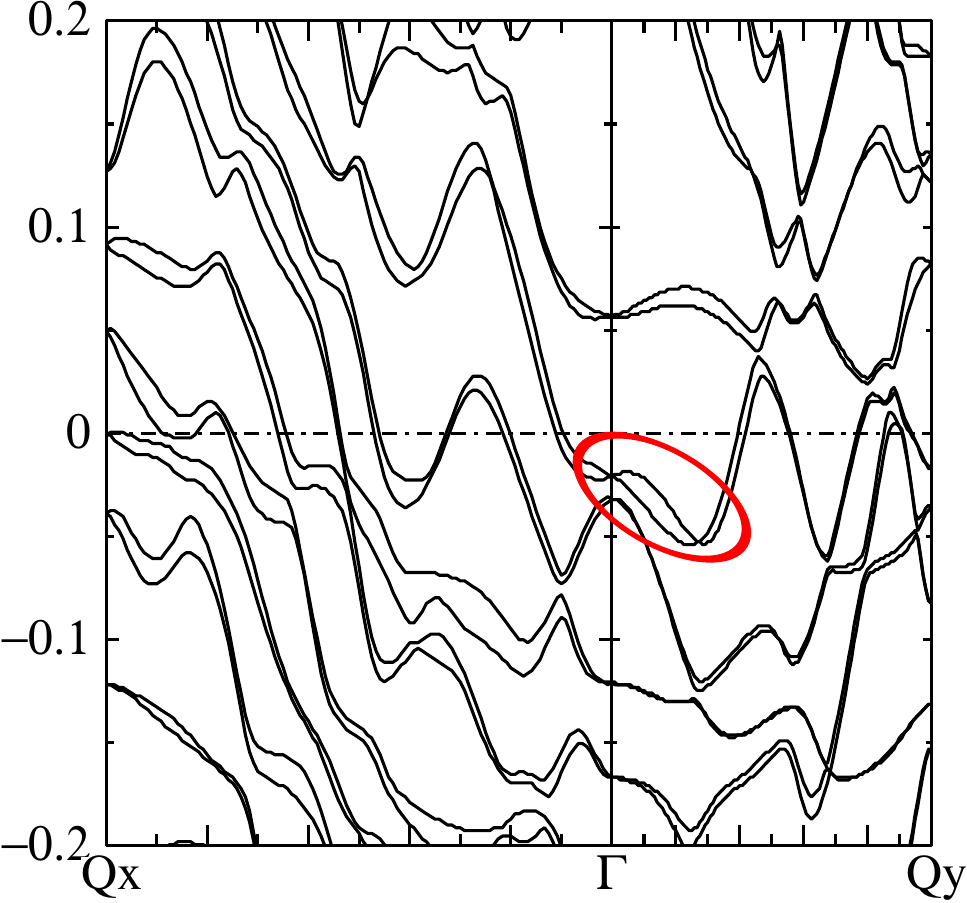}\hspace{5mm}\text{(c)}\includegraphics[width=35mm]{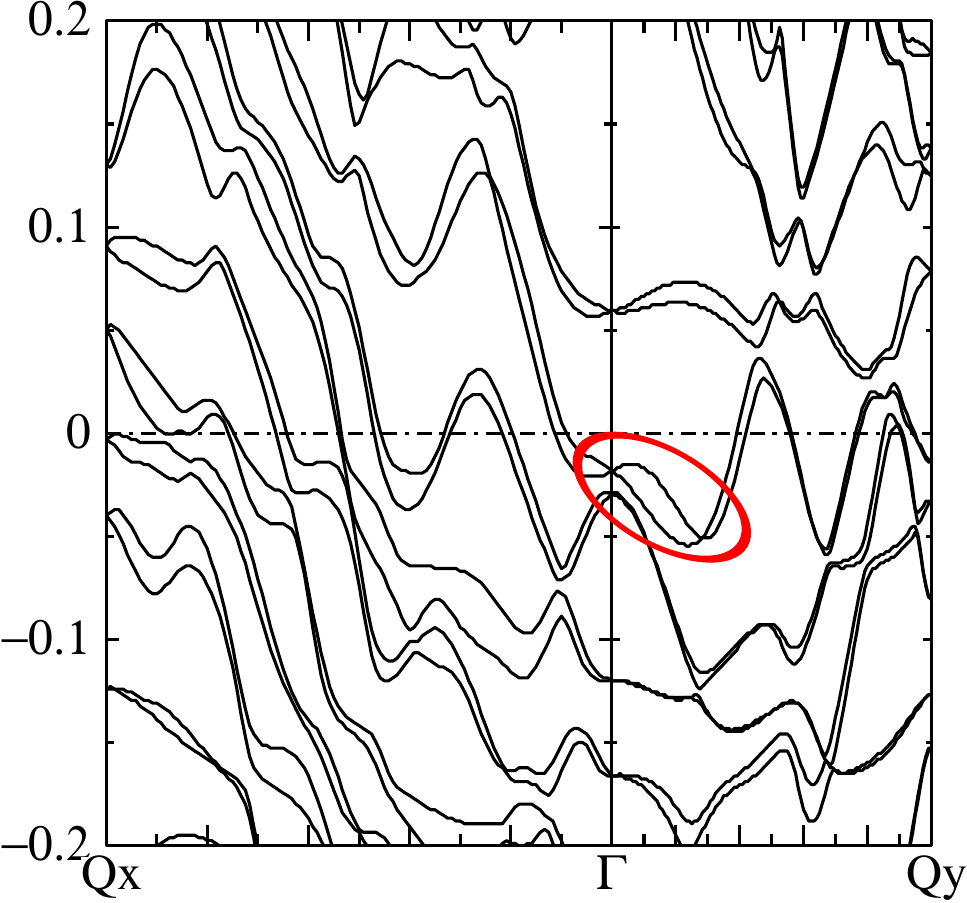}\hspace{5mm}\text{(d)}\includegraphics[width=35mm]{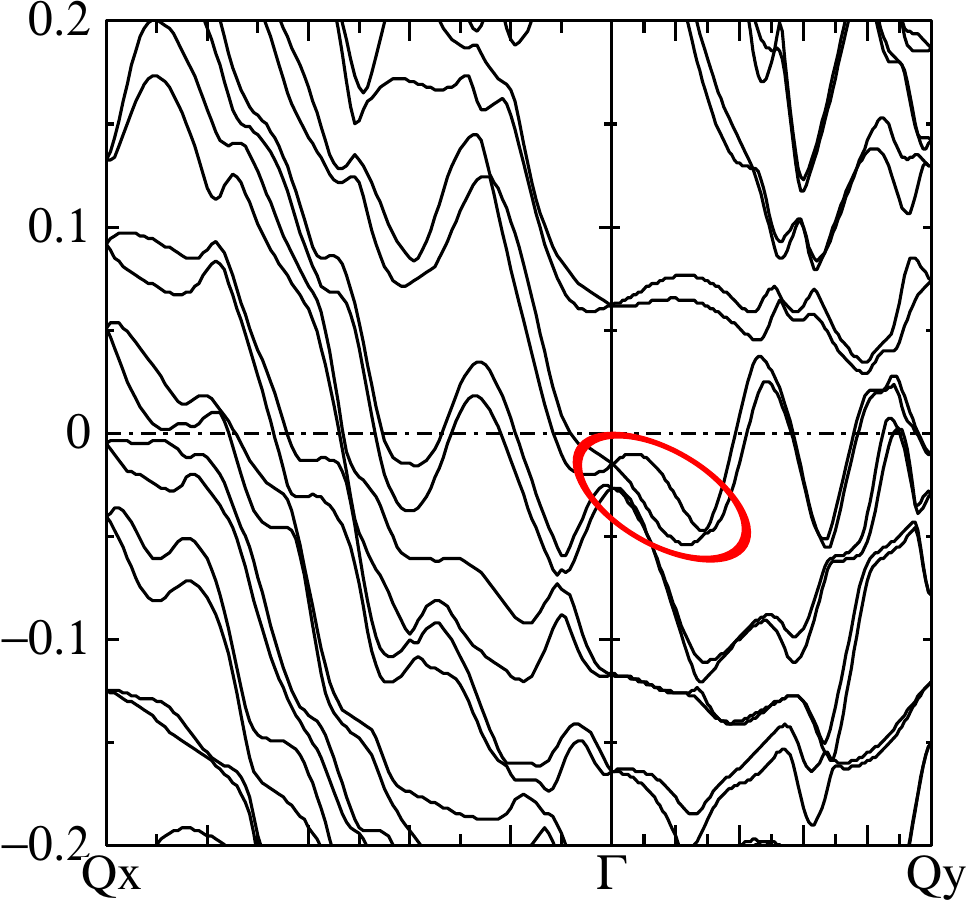}
\caption{\label{fig:bands_InAsAl_rescale}
QS$GW$ band structure of the InAs/Al heterojunction made of 138 atoms,
for different rescaling of the SOC. 
Panel (a), (b), (c) and (d) corresponds to rescaling the original SOC
by $\times 2, \times 3, \times 4$ and $\times 5$ respectively.
Focusing on the bands around the $\Gamma$ point and at $E\sim E_\text{F} - 0.05$,
one can see the spin-split bands (due to SOC) along the $\Gamma-Qy$ direction. The splitting between
the two bands increases with increasing rescaling of the SOC. 
The band encircled in red corresponds to the bottom of the bulk InAs conduction band in the
case of the junction (see Appendix \ref{app:bnds}).
}
\end{figure}

Figure~\ref{fig:spinsplit} shows the energy difference $\Delta E(k)$ of the spin-split bands around the $\Gamma$ point,
bands encircled in red in Fig.~\ref{fig:bands_InAsAl_rescale}, versus the $k$-vector along the $\Gamma-Qy$ direction.
The spin splitting is more important (for small $k$ values away from $\Gamma$) when the SOC rescaling in applied to all In
and As atoms, instead of only on the interface In, As atoms. 
%This is due to some form of cumulative effect.
However, it is clear that, in both cases, the spin splitting $\Delta E(k)$ is linear in $k$ (for $k/Qy<0.1$),
which is most probably the signature of the 2D like character of the corresponding states parallel to the InAs/Al interface
\cite{Luo:2010}.

\begin{figure}
\centering
\text{(a)}\includegraphics[width=70mm]{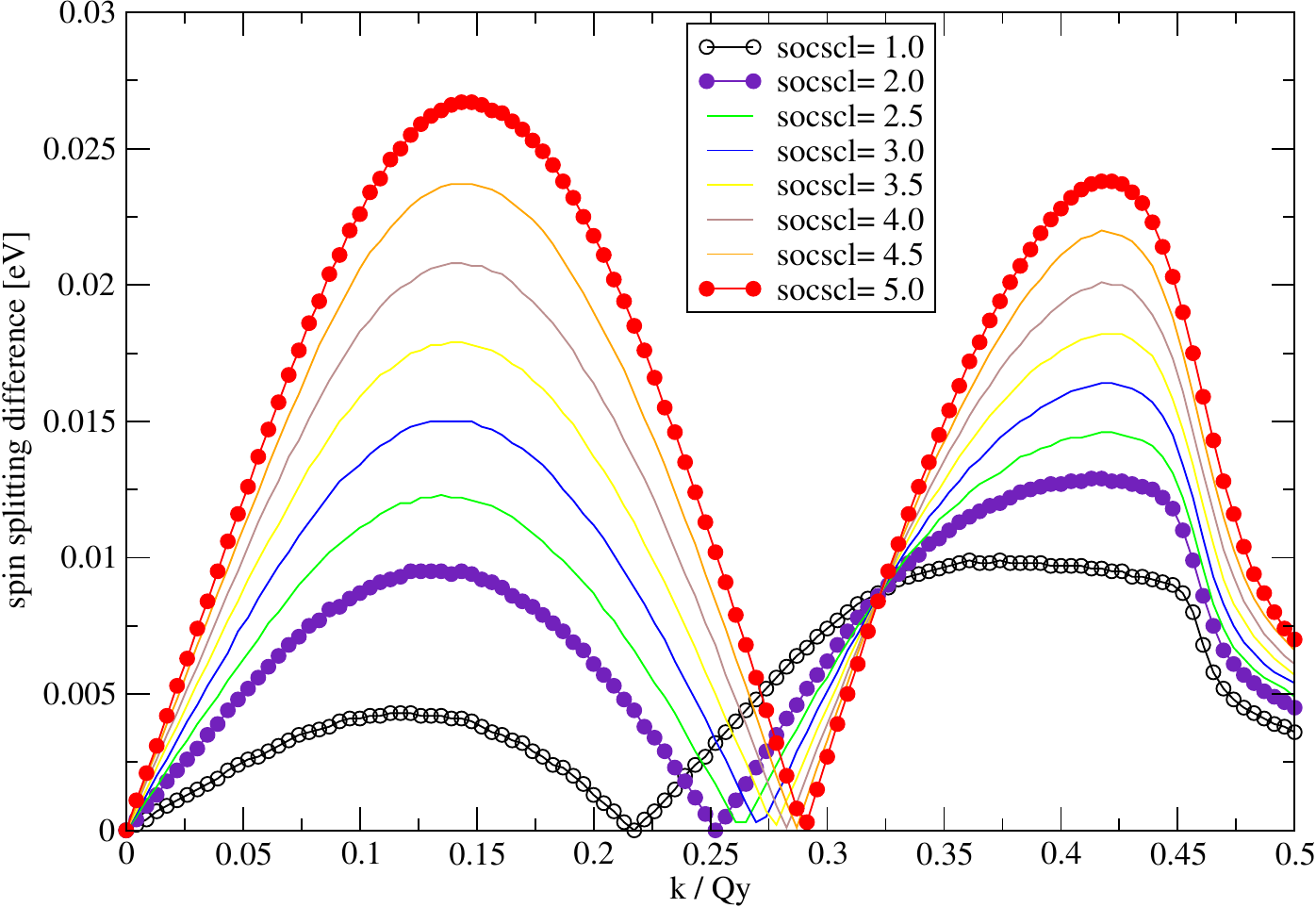}
\text{(b)}\includegraphics[width=70mm]{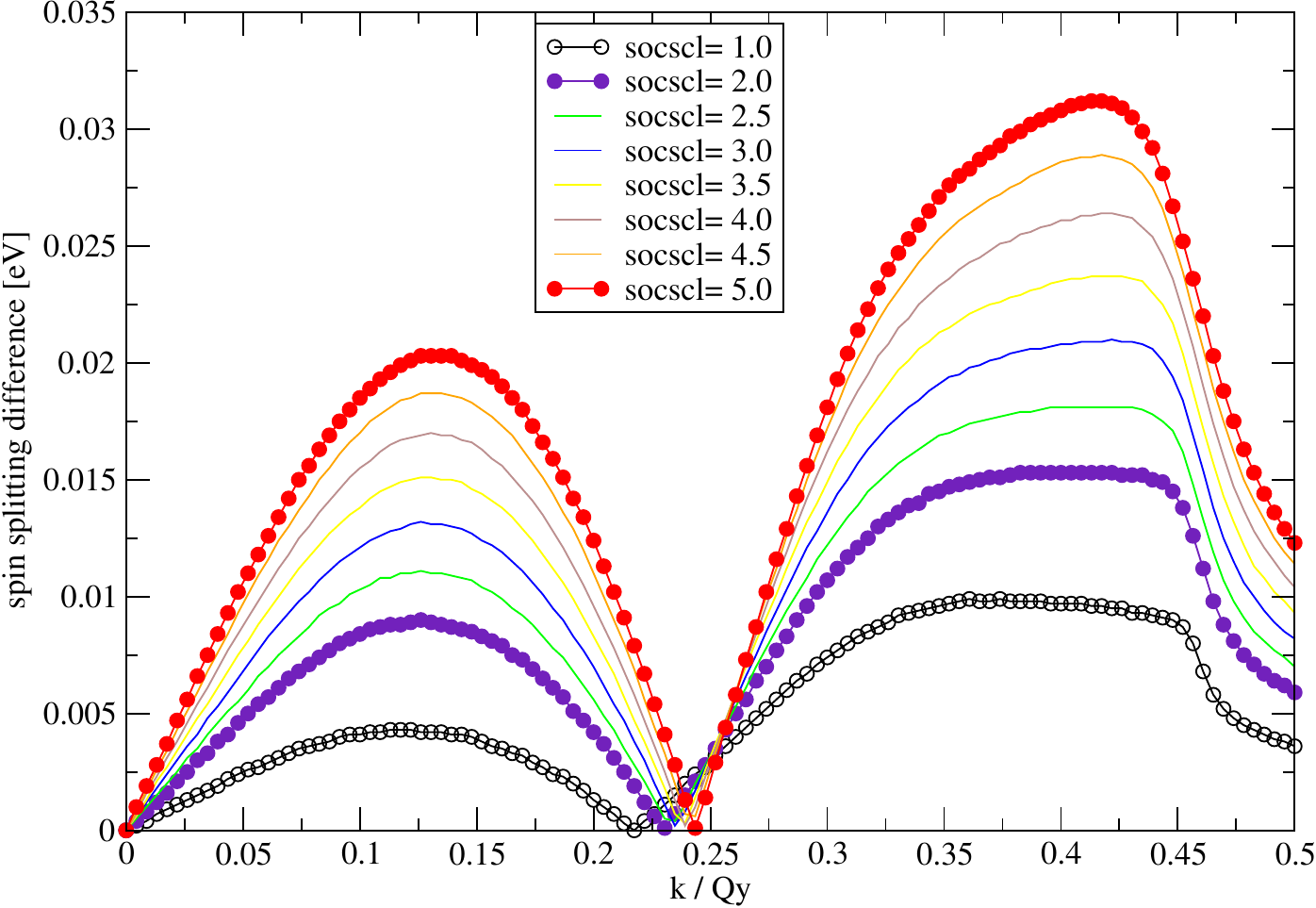}
\text{(c)}\includegraphics[width=50mm]{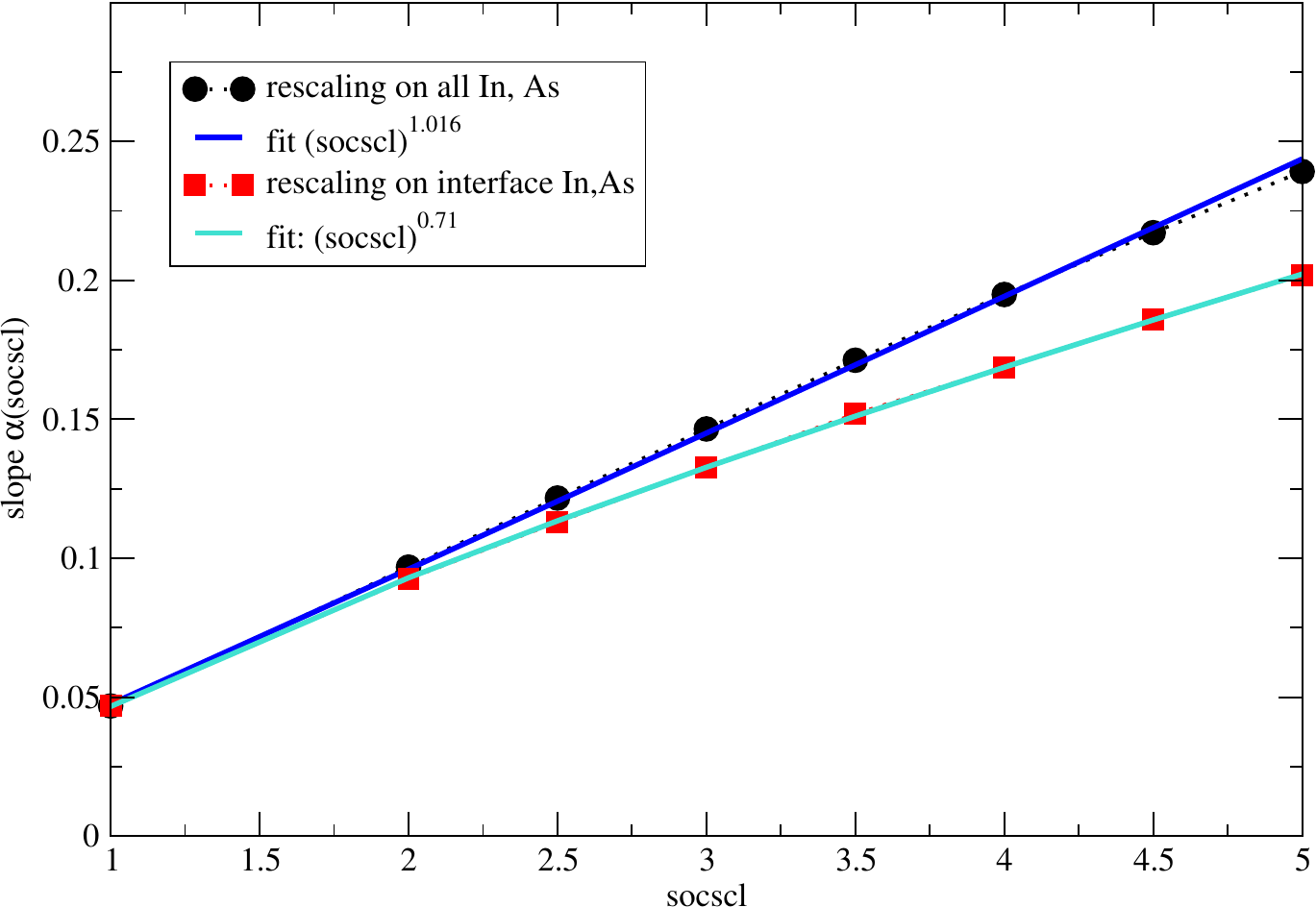}
\caption{\label{fig:spinsplit}
Energy difference $\Delta E(k)$ of the spin-split bands around the $\Gamma$ point in the $\Gamma-Qy$ direction.
The rescaling (soscl) of the spin-orbit coupling is applied to all In and As atoms in panel (a), to only the In and As interface
atoms in panel (b). In both cases, the spin splitting is linear in $k$ for $k/Qy<0.1$ around each band crossing,
i.e. $\Delta E(k) \equiv \alpha k$.
Panel (c): dependence of the slope of $\Delta E(k) \equiv \alpha k$ upon the rescaling soscl of the spin-orbit coupling.
}
\end{figure}

It is also interesting to check how the linear dependence of the spin splitting varies with the rescaling of the SOC.
We determine the slope of the linear relation $\Delta E(k) \equiv \alpha k$ for small $k$ values (for $k/Qy<0.05$).
The dependence of $\alpha(\text{soscl})$ on the SOC rescaling is shown in panel (c) of Figure~\ref{fig:spinsplit}. 
We obtain a linear dependence on SOC rescaling when the rescaling is applied to all In and As atoms, and a sublinear
dependence when the rescaling is only applied to the interface In,As atoms, indicating different screening effects
on the local SOC rescaling.

\section{Conclusion}
\label{sec:ccl}

We have studied the electronic structure of realistic Al/InAs/Al heterojunctions using a combination DFT with hybrid functionals and 
state-of-the-art QS$GW$ calculations.
The InAs/Al heterojunctions we considered are central to superconducting induced properties in InAs and to the design of topological 
quantum computation platforms.
The InAs/Al heterojunctions are described at the atomic level and include atomic relaxations at the InAs/Al interfaces.
Our study confirm the need of well controlled quality of the interfaces to obtain the needed properties of InAs/Al heterojunctions.
The local band alignment (i.e. top of VB, bottom of CB) obtained from QS$GW$ for semiconductor/metal interfaces can be well reproduced 
using dielectric-dependent hybrid functional DFT with the novel metallic estimator which automatically switches off the Fock exchange 
within the bulk of the metal.
The prediction of an accumulation layer for InAs/Al is in agreement with experimental evidence.
The HSE06+DDH method appears to provide an effective approximation to the full QS$GW$ for this system, and open new paths for 
exploring larger interface cells or multiple different interface configurations in relation with experimental devices.
Furthermore, a detailed analysis of the effects of spin-orbit coupling on the spin-splitting of some electronic states show a 
linear scaling in $k$-space. A behaviour most probably related to the two-dimensional nature 
of the interface states.
Our work indicates the possibility of tailoring the properties of the electronic states central to the realisation of topological
computers from the quality of the semiconductor/metal interface.

\begin{acknowledgements}

This work was authored by the National Renewable Energy Laboratory, operated by Alliance for Sustainable Energy, LLC,
for the U.S. Department of Energy (DOE) under Contract No. DE-AC36-08GO28308, funding from Office of Science, Basic
Energy Sciences, Division of Materials. We acknowledge the use of the National Energy Research Scientific Computing
Center, under Contract No. DE-AC02-05CH11231 using NERSC award BES-ERCAP0021783 and we also acknowledge that a portion
of the research was performed using computational resources sponsored by the Department of Energy's Office of Energy
Efficiency and Renewable Energy and located at the National Renewable Energy Laboratory.  For early stages of this
work, HN and MvS acknowledge financial support from Microsoft Station Q via a sponsor agreement between KCL and
Microsoft Research.

HN, DP and MvS acknowledge the Partnership for Advanced Computing in Europe (PRACE) for awarding us access to Juwels Booster and Cluster 
(J\"ulich, Germany).

\end{acknowledgements}

\appendix

\section{Questaal on GPU}
\label{app:gpu}

{Due to the relatively large simulation cell for QS$GW$ standards, together with (i) relatively dense Brillouin zone sampling and 
(ii) high angular momentum cutoffs required, the all-electron, the full frequency QS$GW$ calculations are rather difficult to achieve. 
They were only made practical with the efficient use of new clusters with high-density, high-memory GPU nodes and good interconnect.

Algorithmic improvements avoided most of the filesystem IO. Together with a more flexible memory management, they allowed efficient parallelisation 
across multiple levels of processes and threads enabling various launch configurations. 
Nearly all of the remaining IO was moved to parallel HDF5 maintaining the same file layout independent of the parallelism.

The screened Coulomb potential and the mixed product basis projectors occupy the bulk of the memory available and in the present case could not fit together entirely; 
fortunately the projectors can be generated and used in piecewise fashion with little overhead.

In the GPU context, each device is handled by a thread allowing simple use of multiple devices per process. The threads distribute batches of matrix 
operations across dynamically estimated number of streams depending on dimensions and the available memory. 
In this way host-device transfers and kernel launches are overlapped through asynchronous executions, hiding latency and maximising occupancy and efficiency. 

Most of the compute routines make heavy use of the performance libraries cuBLAS, cuSOLVER, cuFFT and cuSPARSE in this mode. Certain larger matrix operations 
were done collectively with cuSOLVERMg (cu*Mp were not available at the time).

The heaviest step in the computations is the calculation of the full off-diagonal self-energy, it sustained close to 20 PFLOPS on the Juwels-Booster cluster using 288 nodes. 
}

\section{Local DDH functional with metallic correction using localized orbitals}
\label{app:ddh}

{\subsection{DDH overview}

The exchange-correlation energy in HSE is constructed by splitting up the Coulomb interaction in a long and short range part using the error functions
\begin{equation*}
\frac{1}{r} = \frac{\erf(\omega r)}{r} + \frac{\erfc(\omega r)}{r}~,
\end{equation*}
in which $\omega$ is a range separation parameter that determines what is defined as long and short range.
The exchange correlation energy is then split up as:
\begin{align*}
E^{\text{HSE}}_{xc} =& \alpha E^{\text{HF,SR}}_x + (1-\alpha)E^{\text{PBE,SR}}_x(\omega)\\
&\qquad+ E^{\text{PBE,LR}}_x(\omega) + E^{\text{PBE}}_c~.
\end{align*}

The amount of \emph{exact} exchange included is determined by $\alpha$, the exchange fraction. In HSE it is taken to be a constant of 0.25, which is reasonably accurate for medium gap semiconductors but produces some errors for large and small gaps. This is because one can derive that the value of $\alpha$ should be related to the dielectric constant of the material, which is in turn related to the screening. The DDH approach is to create a hybrid functional for which the exchange fraction is determined self-consistently based on the dielectric constant. This is too computationally expensive to calculate, and so an estimator for the dielectric function is used instead (as presented in Refs.~\onlinecite{PhysRevB.83.035119, doi:10.1021/acs.jctc.7b00853}):
\begin{equation*}
\bar{g} = \frac{1}{V}\int d\mathbf{r}~\sqrt{\frac{\nabla\rho(\mathbf{r})}{\rho(\mathbf{r})}}~.
\end{equation*}
The exchange fraction is then related to this estimator by a quartic function:
\begin{equation*}
\alpha_{\text{ddh}} = a_0 + a_4 \bar{g}^4,
\end{equation*}
which we have fitted to the correct experimental band gap for a large set of semiconductors and insulators.

\subsection{Local DDH using localized orbitals}

To study interfaces, it might be that different values of the exchange fraction are needed in different parts of the system. This is why a local estimator is introduced:
\begin{equation*}
\bar{g}(\mathbf{r},\sigma) = \frac{1}{(2\pi\sigma)^{\frac{3}{2}}}\int d\mathbf{r}'~\sqrt{\frac{\nabla\rho(\mathbf{r}')}{\rho(\mathbf{r}')}}\exp\left(-\frac{|\mathbf{r}-\mathbf{r}'|^2}{2\sigma}\right)~,
\end{equation*}
from which we can calculate an exchange fraction field
\begin{equation*}
a(\mathbf{r},\sigma) = a_0 + a_4 \bar{g}(\mathbf{r},\sigma)^4.
\end{equation*}
 In Ref.~\onlinecite{doi:10.1021/acs.jctc.7b00853} this is used in the calculation of the integrals, so that the exchange matrix becomes
\begin{equation*}
X_{ij} = \sum_{kl} V_{ik;jl}D_{kl}~,
\end{equation*}
where the integrals are defined as:
\begin{equation*}
V_{ik;jl} = \int d\mathbf{r}d\mathbf{r}' \phi_i(\mathbf{r})\phi_k(\mathbf{r})\alpha(\mathbf{r},\mathbf{r}';\sigma)K(|\mathbf{r}-\mathbf{r}'|)\phi_j(\mathbf{r}')\phi_l(\mathbf{r}')~,
\end{equation*}
with
\begin{equation*}
\alpha(\mathbf{r}, \mathbf{r}';\sigma)= \sqrt{a(\mathbf{r},\sigma)a(\mathbf{r}',\sigma)}
\end{equation*}
and $K$ the short range Coulomb kernel.

Instead of recalculating the integrals we use the fact that in QuantumATK a resolution of identity (RI)~\cite{Ren_2012} approach is used to calculate the Coulomb integrals:
\begin{equation*}
V_{ik;jl}\approx C^\mu_{ik}V_{\mu\nu}C^\nu_{jl}~,
\end{equation*}
with
{the introduction of an auxiliary basis $P_\mu(\mathbf{r})$ such that}
\[
\phi_i(\mathbf{r})\phi_j(\mathbf{r}) = \sum_\mu C^\mu_{ij}P_\mu(\mathbf{r})~.
\]
We assume that the auxiliary basis coefficients, $C^\mu_{ij}$ are unaffected, and only the integrals between the auxiliary basis functions change:
\begin{align*}
V_{\mu\nu}=&\int d\mathbf{r}d\mathbf{r}' \tilde{\phi}_\mu(\mathbf{r})\alpha(\mathbf{r},\mathbf{r}';\sigma)K(|\mathbf{r}-\mathbf{r}'|)\tilde{\phi}_\nu(\mathbf{r}')\\
\approx&\sqrt{\bar{a}_\mu\bar{a}_\nu}\int d\mathbf{r}d\mathbf{r}' \tilde{\phi}_\mu(\mathbf{r})K(|\mathbf{r}-\mathbf{r}'|)\tilde{\phi}_\nu(\mathbf{r}')~,
\end{align*}
in which we have taken a Gaussian average of the $a$ coefficients around the center they are located on:
\begin{equation}
\bar{a}_\mu(\sigma, \tau) = \int d\mathbf{r} a(\mathbf{r},\sigma)\exp\left(-\frac{|\mathbf{r}-\mathbf{r}_\mu|^2}{2\tau}\right)~.
\label{average_around_center}
\end{equation}
The approximation being made here is that the local estimator is approximately constant or at least slowly varying over the region where a single auxiliary basis function has support, which is on a center.

\subsection{Metallic correction to local DDH}

For metals there is perfect screening, and we would expect the exchange fraction to go down to zero. Unfortunately the DDH method doesn't reproduce this behaviour. This is why we introduce a second \emph{metallic} estimator. At every step of the self-consistent loop we calculate the Fermi level density matrix
\[
\mathcal{F}_{ij} = \sum_\mathbf{k} \sum_n \exp{\left[-\frac{(\epsilon_F - \epsilon_{\mathbf{k},n})^2}{2\sigma_F}\right]} \braket{\phi_i}{\psi_{\mathbf{k}n}}\braket{\psi_{\mathbf{k}n}}{\phi_j}~,
\]
where $\sigma_F$ is the Fermi level broadening, chosen to be 0.001~eV.
This is used to calculate the Fermi level density
\[
f(\mathbf{r}) = \sum_{ij}\mathcal{F}_{ij}\phi_i(\mathbf{r})\phi_j(\mathbf{r})~.
\]
We then define the following metallic estimator function:
\begin{align*}
    M(\mathbf{r}) &= 1 \quad\text{if} \quad f(\mathbf{r}) < c_\mu \\
    M(\mathbf{r}) &= 0 \quad\text{if} \quad f(\mathbf{r}) \geq c_\mu
\end{align*}
where $c_\mu$ is a cutoff parameter we have chosen to be 0.0003. This function is convoluted with a Gaussian to get a smooth metallic estimator function:
\[
m(\mathbf{r}) = \int d\mathbf{r}' M(\mathbf{r}') \exp\left(-\frac{|\mathbf{r}-\mathbf{r}'|^2}{2\sigma_\mu}\right)~,
\]
where the width of the Gaussian is chosen to be 1~{\AA}. 
We then multiply the metallic estimator with the function $a(\mathbf{r},\sigma)$:
\[
a_m(\mathbf{r},\sigma) = m(\mathbf{r}) a(\mathbf{r},\sigma)~,
\]
before the averaging around a center is performed in Eq.~(\ref{average_around_center}).
}

\pagebreak
\pagebreak
\newpage
\pagebreak

\onecolumngrid 
\section{Supplementary Material}
\label{app:extra}

\newpage
\subsection{Local density of states}
\label{app:ldos}

\begin{figure}
\centering
\text{(a)}\includegraphics[width=80mm]{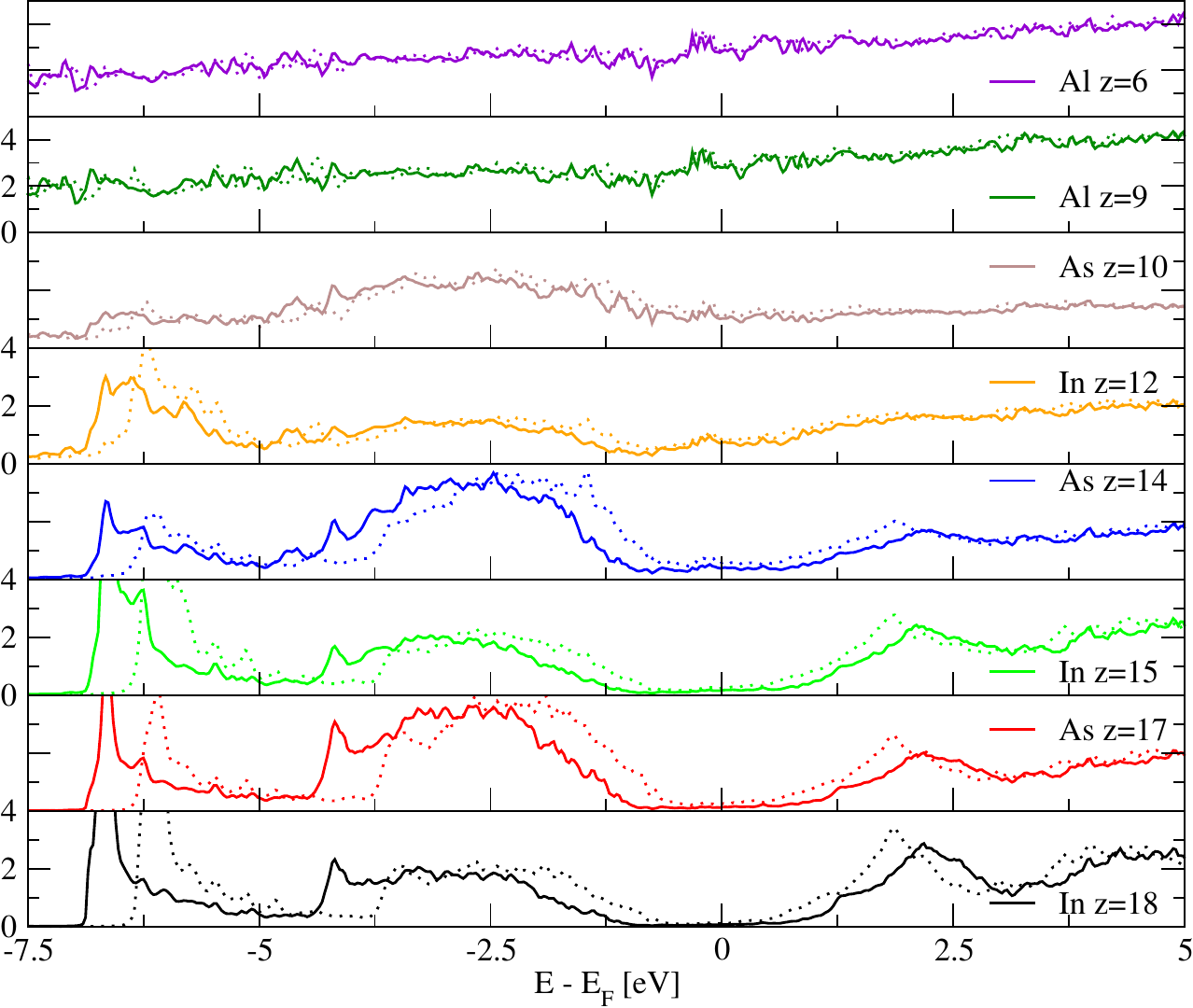}
\text{(b)}\includegraphics[width=80mm]{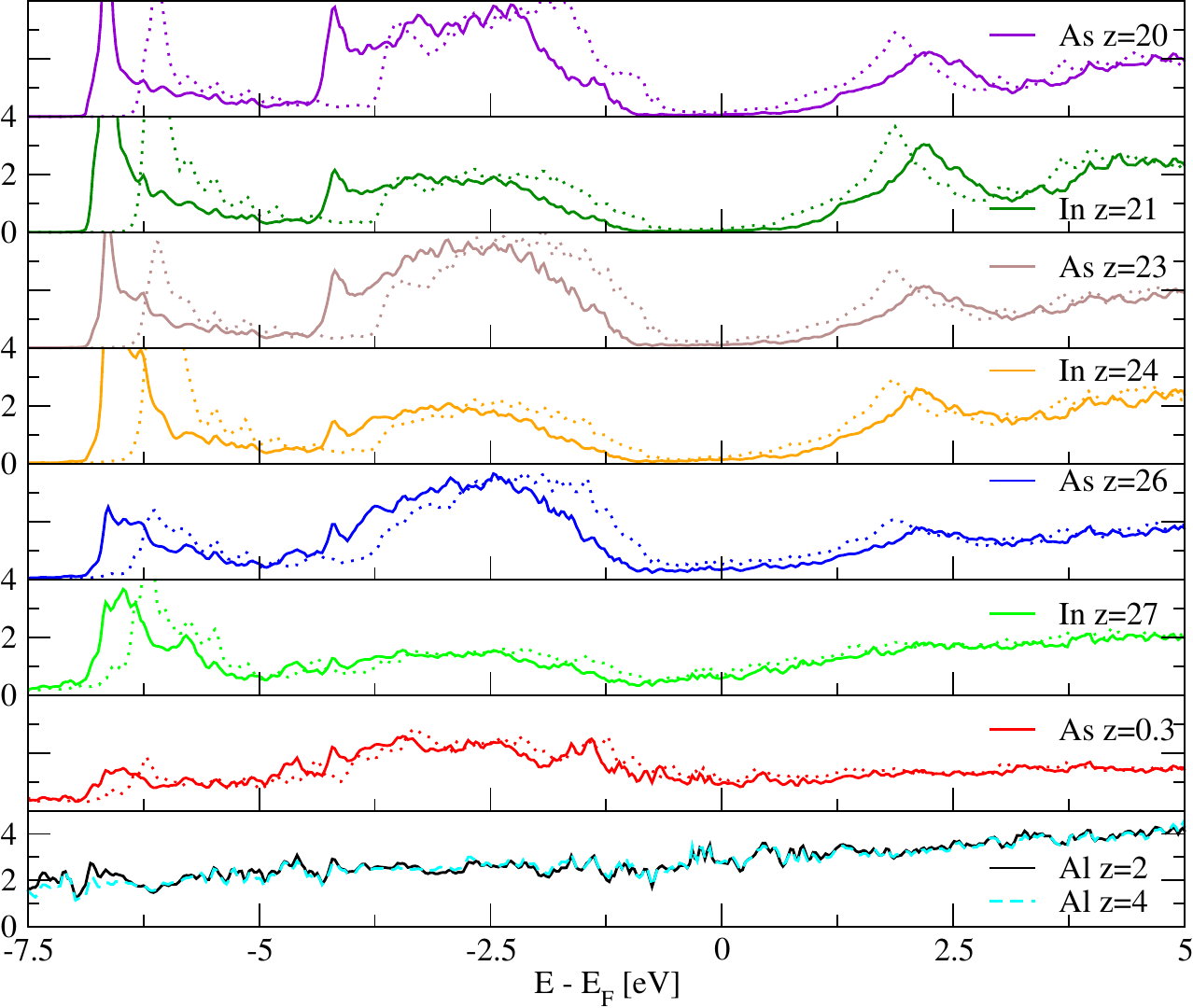}
\text{(c)}\includegraphics[width=50mm]{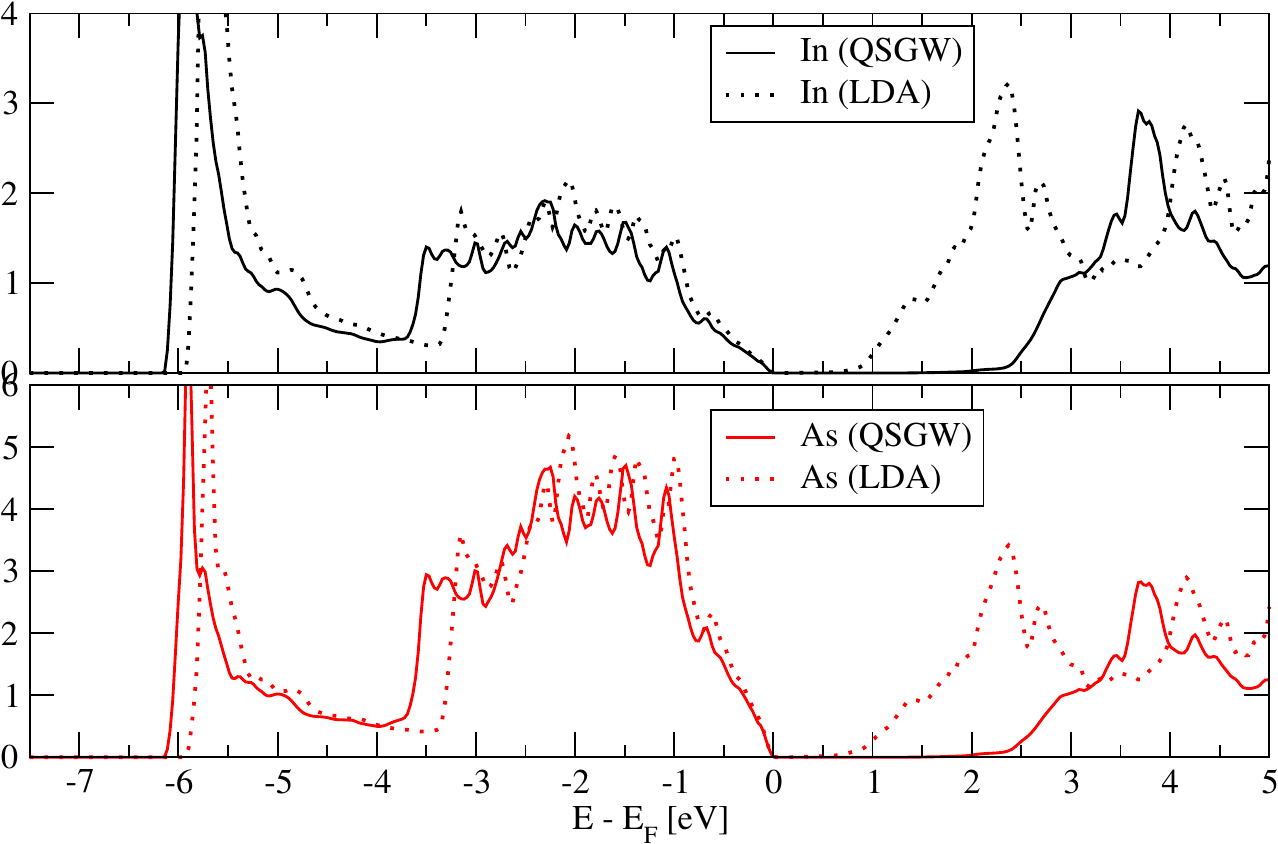}\hspace{10mm}\text{(d)}\includegraphics[height=40mm]{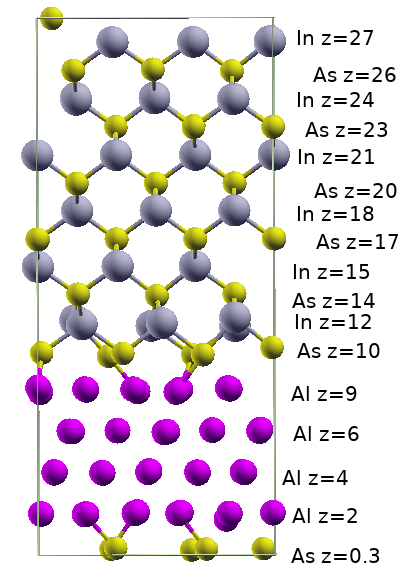}
\caption{\label{fig:localDOS}
Local density of states LDOS per atomic layers in the InAs/Al heterojunction, panels (a) and (b),
obtained from QS$GW$ (solid lines) and LDA (dotted lines) calculations.
Panel (c) shows the corresponding bulk LDOS,
panel (d) shows the labelling of the atomic layers in the $z$-direction.
The LDOS of the central In and As atomic layers, labelled $z=18, 20, 21$, are similar
to the bulk LDOS.
The LDOS of the In and As atomic layers acquires a stronger admixture with the
Al states, the closer the layers are to the InAs/Al interfaces.
Note that in the bulk LDOS, the Fermi energy $E_\text{F}$ is located at the top of the
valence band as a convention for any semiconductors at zero temperature. For the heterojunction, 
the position of $E_\text{F}$ is governed by the metallic states of the Al slab.}
\end{figure}

\newpage
\subsection{Transition bulk to heterojunction}

\begin{figure}
\centering
\includegraphics[width=80mm]{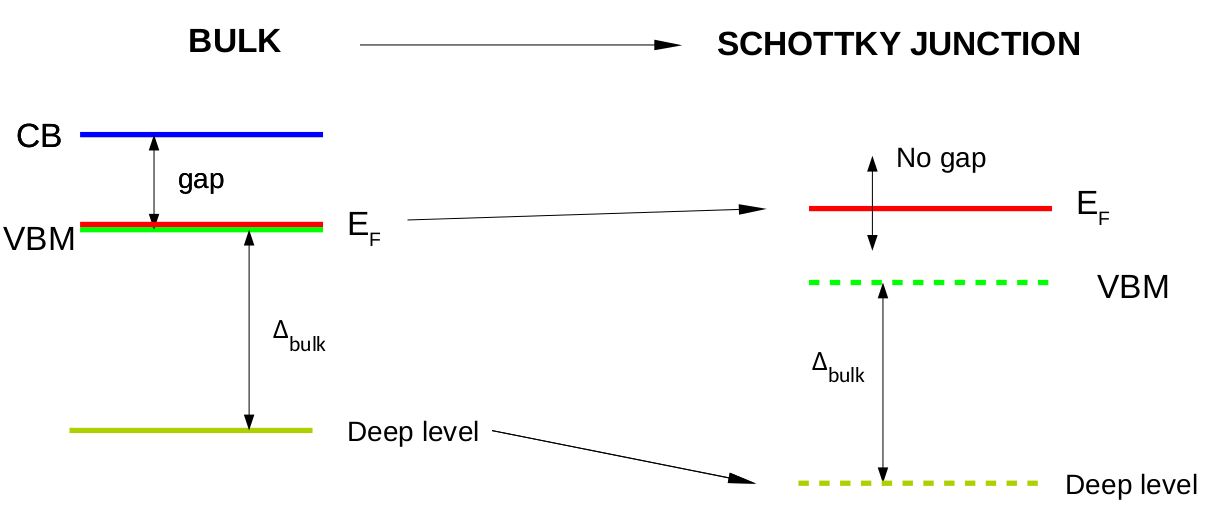}
\caption{\label{fig:deeplevelshift}
Schematic representation of the modifications of the In and As deep electronic
levels upon changing atoms from a bulk configuration to a InAs/Al heterojunction
configuration.
}
\end{figure}

\newpage
\subsection{Band structure}
\label{app:bnds}

\begin{figure}
\centering
\includegraphics[width=50mm]{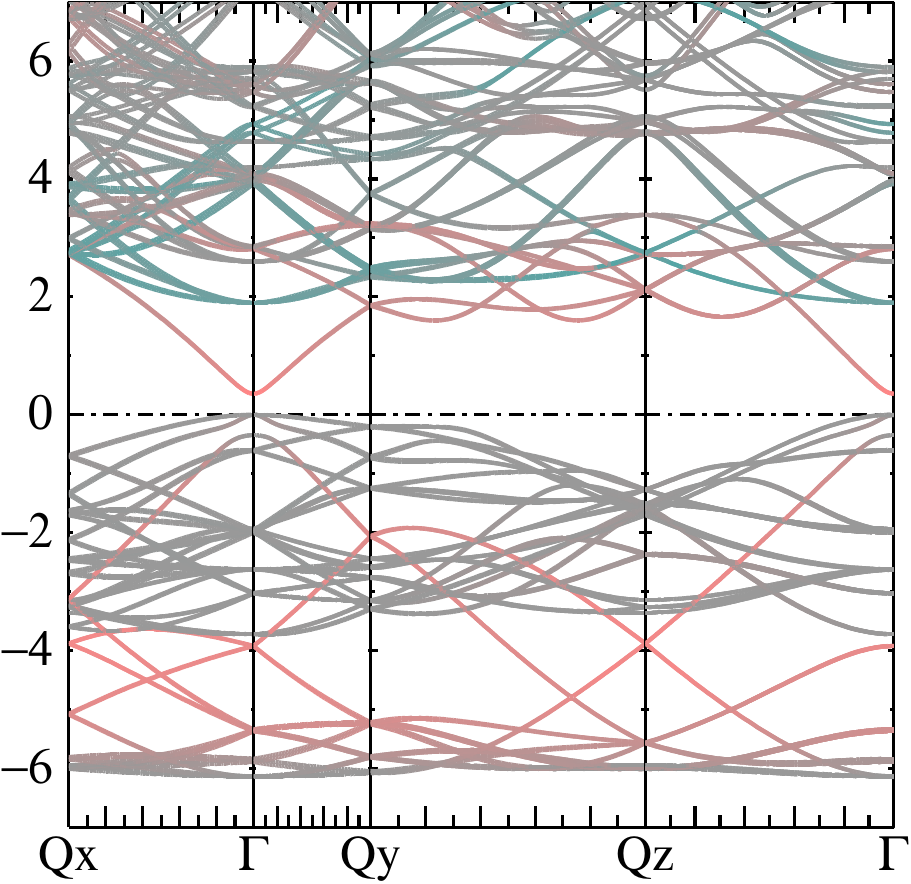}\includegraphics[width=50mm]{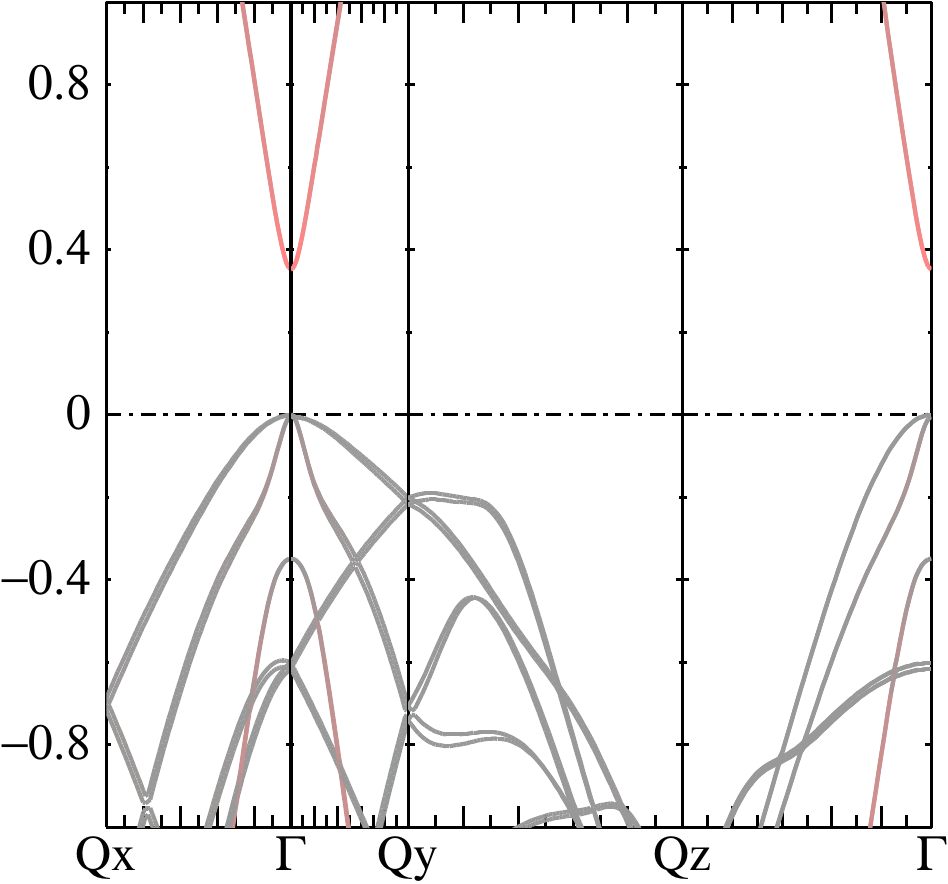}
\caption{\label{fig:bands_bulkInAs_exten}
QS$GW$ Bulk band structure of InAs. Energies are given in eV.
For allowing direct comparisons between bulk
and heterojunction systems, the bulk bands have been calculated for a cell
having the same lattice vectors $u_{1,2}$ in the (001) plane as for the
InAs/Al heterojunction. $Qx,y,z$ represents the (x,y,z)-direction in the reciprocal
space. 
The color scheme represent the weight of the As $s$-orbital in the
bands: red: large weight of As $s$-orbital, grey/black: no As $s$-orbital
weight. One can see that the bottom of the conduction band is mostly consisting
of As $s$-states.
This color scheme will be useful to identify the ``bulk InAs conduction band''
in the band structure of the InAs/Al heterojunction.
}
\end{figure}

\newpage
\begin{figure}
\centering
\text{(a)}\includegraphics[width=50mm]{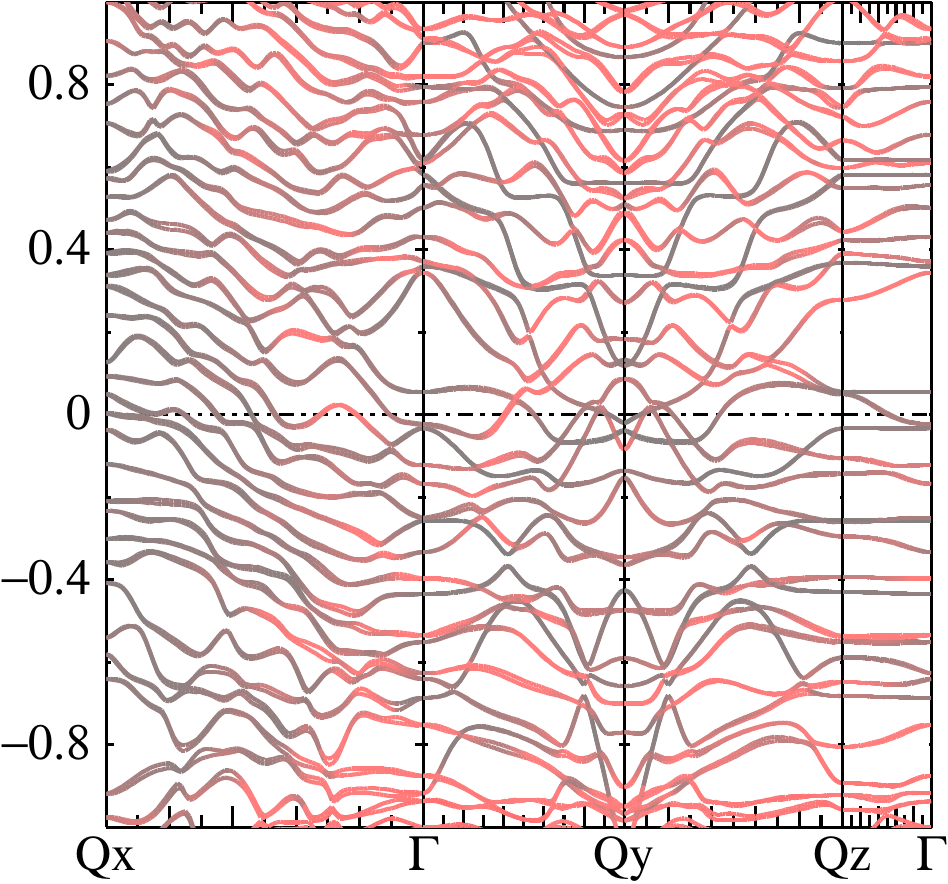}\hspace{5mm}\text{(b)}\includegraphics[width=50mm]{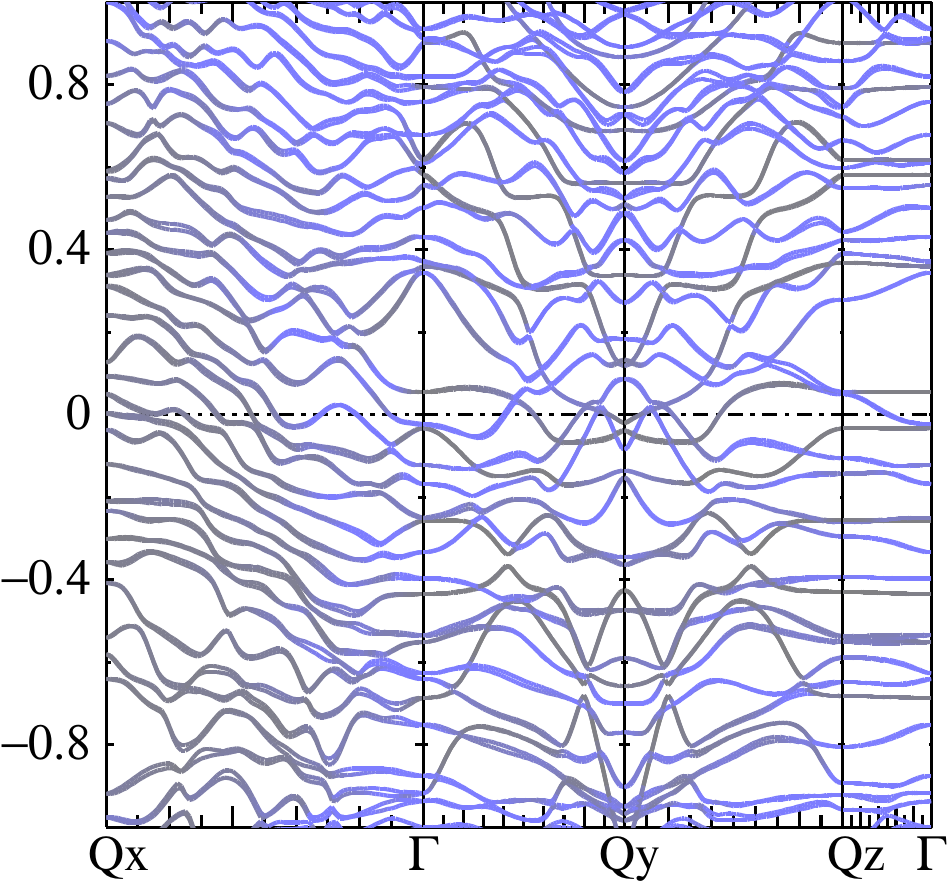}\hspace{5mm}\text{(c)}\includegraphics[width=50mm]{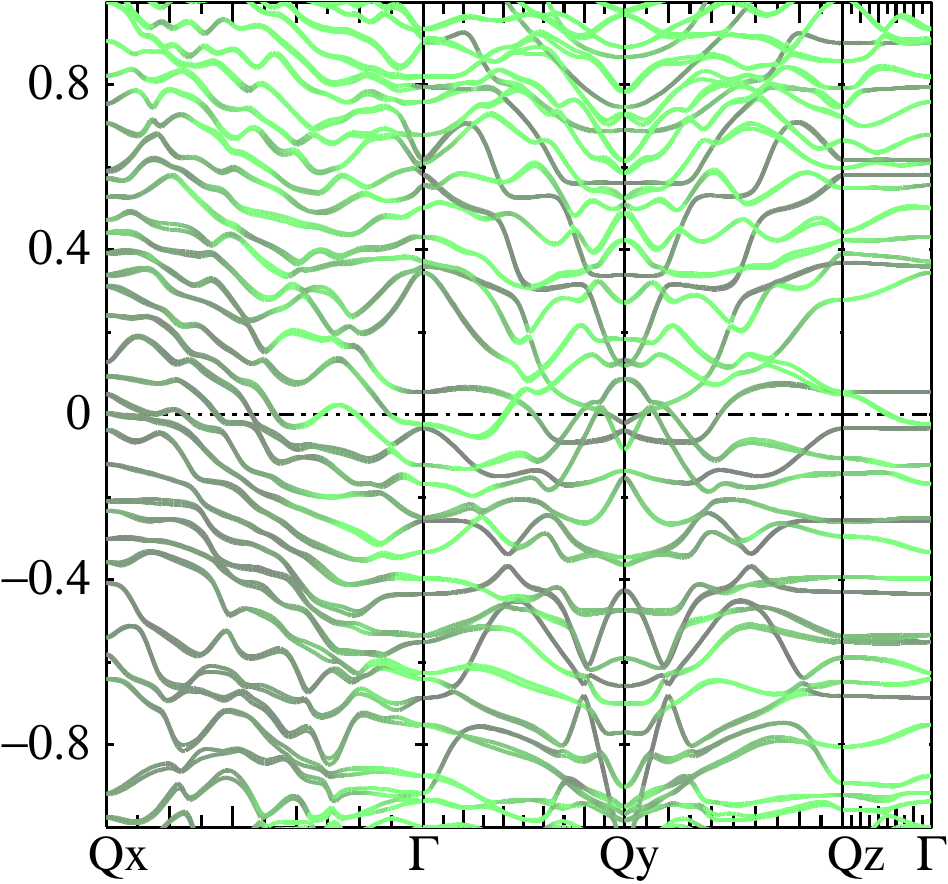}
\caption{\label{fig:bands_InAsAl}
QS$GW$ band structure of the InAs/Al heterojunction made of 138 atoms. Energies are given in eV.
The band structure look much more complex than for bulk InAs due to the large size of the system, the presence of the Al metallic states
and the relaxation of the InAs/Al interfaces. There is clearly no band gap anymore in the heterojunction.
The bands in panels (a), (b), (c) are colored in red, blue, green respectively according to the projection weight 
of the states onto the bulk-like atomic layers of As $z=20$, In $z=18$, In $z=21$. See panel (d) in Fig.~\ref{fig:localDOS}
for the $z$-direction labelling. There is a strong mixing of all As and In orbitals in the bands, as well as mixing with Al orbitals.
Note the two flat bands around $E_\text{F}$ along the $\Gamma-Qz$ direction implying the existence of localized ``interface'' states. This states
are however delocalised in the ($xy$) planes of the InAs/Al interfaces. A careful analysis of the composition of these states (not shown
here) reveal that they consist mostly of orbitals of the As $z=10$ and Al $z=9$ (and of the As $z=0.3$ and Al $z=2$) atomic layers.
}
\end{figure}

\newpage
\begin{figure}
\centering
\includegraphics[width=50mm]{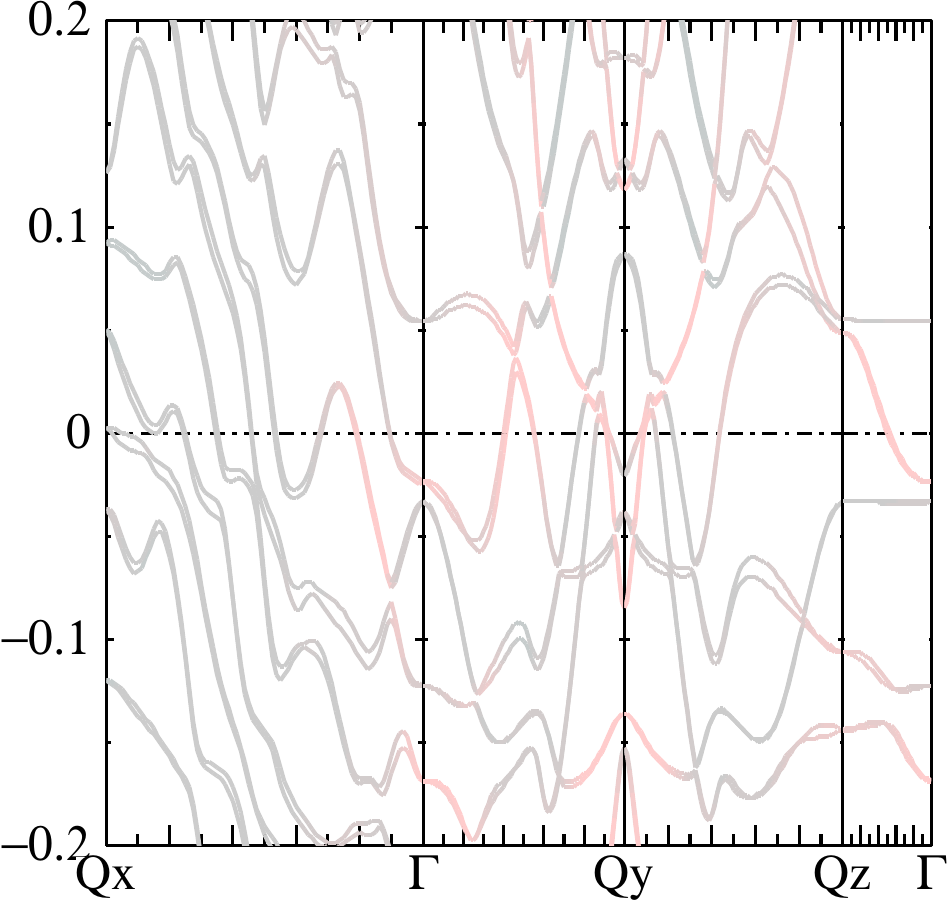}
\caption{\label{fig:bands_InAsAl_zoom}
QS$GW$ band structure of the InAs/Al heterojunction made of 138 atoms. Energies are given in eV.
Zoom in energy of the bands shown in Fig.~(\ref{fig:bands_InAsAl}).
The bands are colored in red according to the projection weight 
of the states onto the $s$-orbitals of the bulk-like atomic layers of As $z=20$.
In Fig.~(\ref{fig:bands_bulkInAs_exten}), we have seen that the bottom of the bulk InAs
conduction band, around the $\Gamma$ point, is mostly of As $s$-character.
For the InAs/Al heterojunction, we can see that some bands around 
the $\Gamma$ point, still keep a non-negligible weight of the bulk-like As $s$-orbitals;
more specially the bands $\sim$ 50 meV below $E_\text{F}$.
This energy position below  $E_\text{F}$ corresponds well with our estimate of the energy shift
of the bulk-like conduction band minimum shown in Fig.~(\ref{fig:localVBM}).
We therefore consider that these bands are the equivalent of the bottom of the bulk-like conduction band 
of InAs which strongly couples to the Al states in the InAs/Al heterojunction.
}
\end{figure}

\pagebreak
\begin{figure}
%\centering
\text{(a)}\includegraphics[width=35mm]{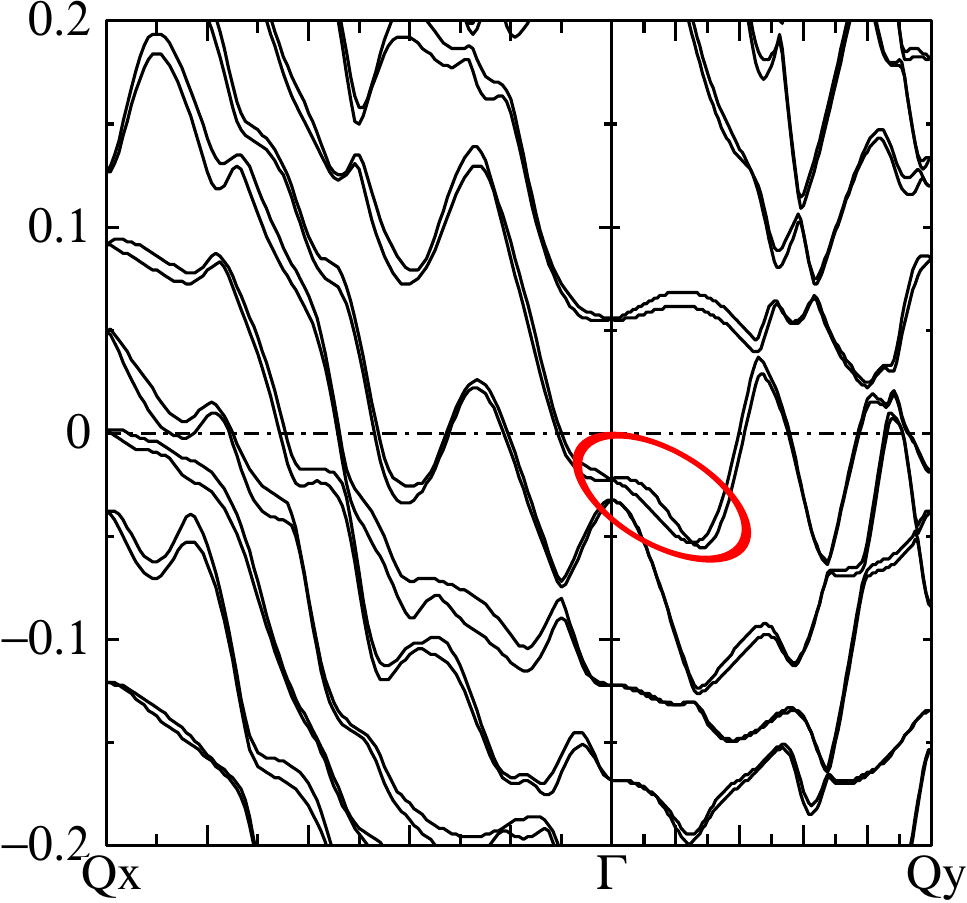}\hspace{5mm}\text{(b)}\includegraphics[width=35mm]{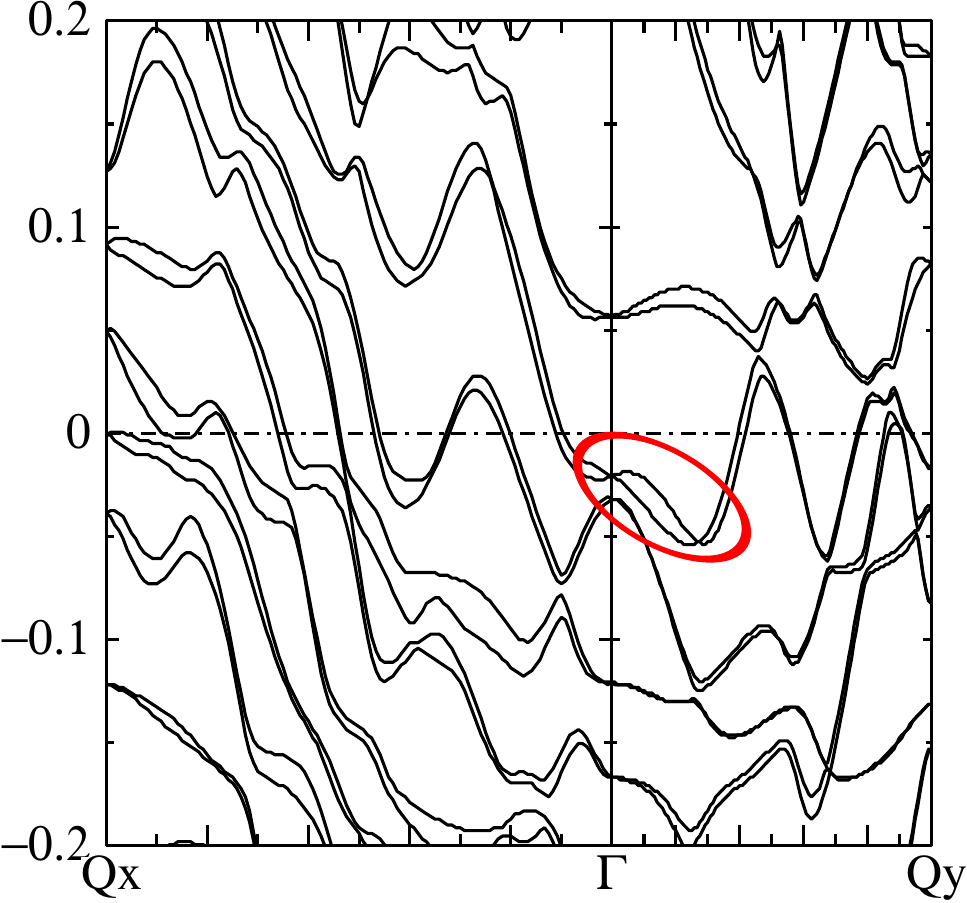}\hspace{5mm}\text{(c)}\includegraphics[width=35mm]{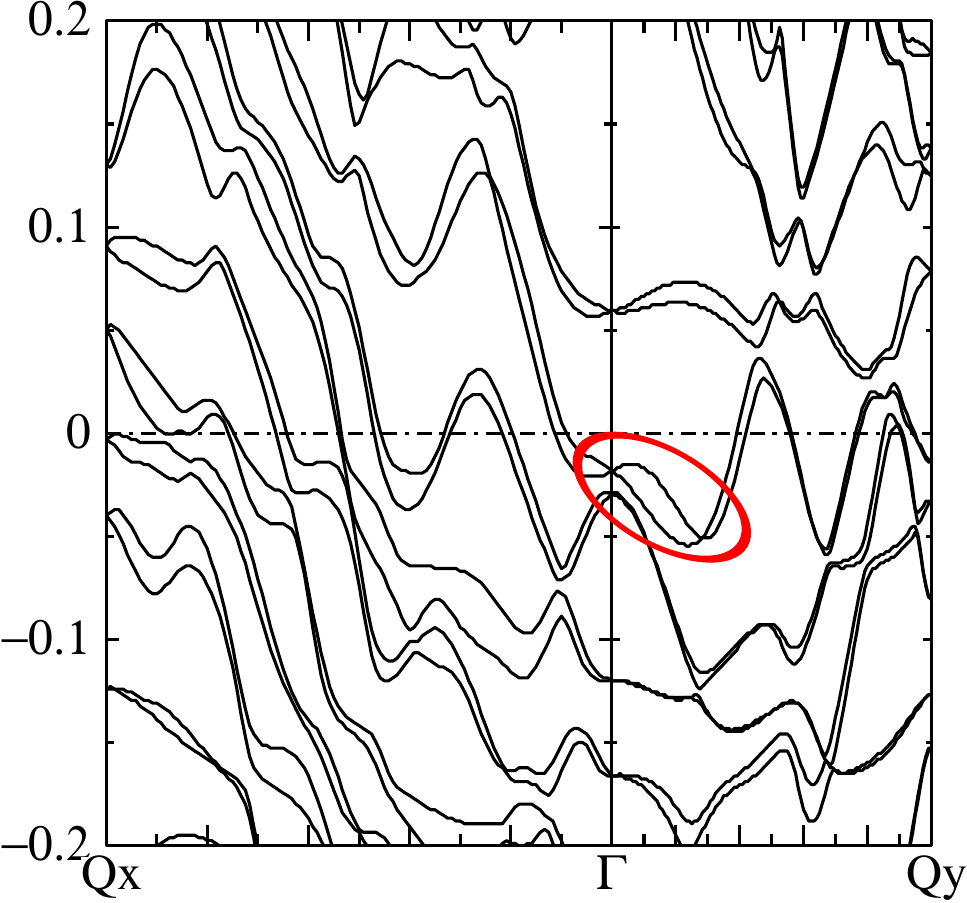}\hspace{5mm}\text{(d)}\includegraphics[width=35mm]{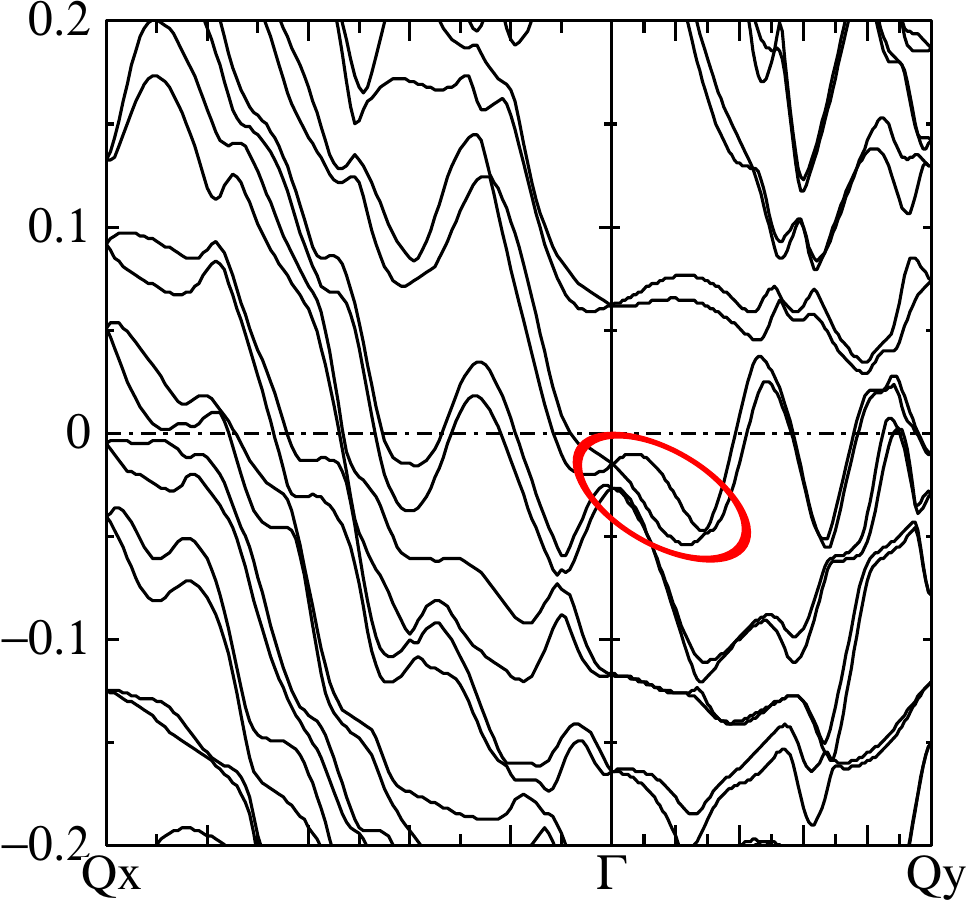}
%
%\vspace{-10mm}
%\begin{tikzpicture}
%    \draw[red] [xshift= 0mm, yshift= 15mm] (0.0,0) ellipse (3mm and 2mm) ;
%    \draw[red] [xshift= 5mm, yshift= 15mm] (0,0) ellipse (3mm and 2mm);
%    \draw[red] [xshift= 0mm, yshift= 15mm] (12.,0) ellipse (3mm and 2mm);
%    \draw[red] [xshift= 0mm, yshift= 15mm] (16.5,0) ellipse (3mm and 2mm);
%\end{tikzpicture}
%
\caption{\label{app:fig:bands_InAsAl_rescale}
QS$GW$ band structure of the InAs/Al heterojunction made of 138 atoms,
for different rescaling of the SOC. 
Panel (a), (b), (c) and (d) corresponds to rescaling the original SOC
by $\times 2, \times 3, \times 4$ and $\times 5$ respectively.
Focussing on the bands around the $\Gamma$ point and at $E\sim E_\text{F} - 0.05$,
one can see the spin-split bands (due to SOC) along the $\Gamma-Qy$ direction. The splitting between
the two bands increases with increasing rescaling of the SOC, as expected. }
\end{figure}

\end{document}